\documentclass[10pt]{article}
\usepackage{amsmath, latexsym, amssymb, amscd, amsfonts, amsthm, epsfig}
\usepackage[english]{babel}

\DeclareFixedFont{\sfracFont}{U}{euf}{b}{n}{7pt}
\newtheoremstyle{mydefi}
  {15pt}
  {15pt}
  {}
  {}
  {\bfseries}
  {:}
  {.5em}
  {}
\newtheoremstyle{mytheo}
  {15pt}
  {15pt}
  {\slshape}
  {}
  {\bfseries}
  {:}
  {.5em}
  {}

\theoremstyle{mytheo}
\newtheorem{stel}{Theorem}[section]
\newtheorem{lem}[stel]{Lemma}

\theoremstyle{mydefi}
\newtheorem{de}[stel]{Definition}

\newcommand{\BB}[1]{\mathbb{#1}}

\newcommand{\half}{{\textstyle \frac{1}{2}}}
\newcommand{\ten}{\otimes}

\newcommand{\CC}{\mathcal{C}}

\newcommand{\FC}{\mathcal{F}}

\newcommand{\SC}{\mathcal{S}}

\newcommand{\x}{\hbox{\sfracFont X}}

\begin{document}
\title{Optimal Pointers for Joint Measurement of $\sigma_x$ and $\sigma_z$
via Homodyne Detection}
\author{Bas Janssens$^*$ and Luc Bouten$^\dagger$\\ \\
\normalsize{$^*$Department of Mathematics, Radboud University Nijmegen,}\\
\normalsize{ Toernooiveld 1, 6525 ED, Nijmegen, The Netherlands}\\ \\
\normalsize{$^\dagger$Physical Measurement and Control 266-33, California Institute 
of Technology,}\\ \normalsize{1200 E.\ California Blvd., Pasadena, CA 
91125, USA}}
\date{}
\maketitle

\vspace{8mm}

\begin{abstract}  
We study a model of a qubit in 
interaction with the electromagnetic field. 
By means of homodyne detection, 
the field-quadrature $A_t+A_t^*$ is 
observed continuously in time.
Due to the interaction, 
information about the initial state of the qubit is transferred into 
the field, thus influencing the homodyne measurement results.  
We construct random variables
(pointers) on the probability space of homodyne measurement outcomes having 
distributions close to the initial distributions of $\sigma_x$ and 
$\sigma_z$. Using variational calculus, we find the pointers that 
are optimal. These optimal pointers are very close to 
hitting the bound imposed by Heisenberg's uncertainty relation 
on joint measurement of two non-commuting observables. We close 
the paper by giving the probability densities of the pointers.          
\end{abstract}

\section{Introduction}\label{sec intro}

The implementation of quantum filtering and 
control \cite{Bel88} in recent experiments \cite{AASDM02}, \cite{GSM04} 
has brought new interest to the field of continuous time 
measurement of quantum systems \cite{Dav69}, \cite{Dav76}, \cite{Bel88}, 
\cite{Bel92b}, \cite{Car93}, \cite{WiM93a}, \cite{BGM04}.
In particular, homodyne detection has played a considerable role in this
development \cite{Car93}. 
In this paper, we aim to gain insight into the transfer of information 
about the initial state of a qubit from this qubit, a two-level atom, 
to the homodyne photocurrent, which is observed in actual experiments. 
Our goal is to perform 
a joint measurement of two non-commuting observables in the 
initial system. In order to achieve this, we construct random variables 
(pointers) on the space of possible homodyne measurement results, having 
distributions close (in a sense to be defined) to the 
distributions of these observables in the initial state.  

The problem of joint measurement of non-commuting 
observables has been studied by several authors before, see 
\cite{vNe32}, \cite{Dav76}, \cite{Hol82} and the references 
therein. As a measure for the quality of an unbiased measurement, we use the 
difference between the variance of the pointer in the final state and the 
variance of the observable in the initial state, evaluated in  
the worst case initial state \cite{Jan05}. In other words, the quality of 
measurement is given in terms of the worst case 
added variance. These worst case added variances for two 
pointers, corresponding to two non-commuting observables 
of the initial system, satisfy a Heisenberg-like relation 
that bounds how well their joint measurement can be 
performed \cite{Jan05}.   

The paper concentrates on the example of a qubit 
coupled to the quantized electromagnetic field. We study 
this system in the weak coupling limit \cite{Gou05}, i.e.\ 
the interaction between qubit and field is governed by a 
quantum stochastic differential equation in the sense of 
Hudson and Par\-tha\-sa\-ra\-thy \cite{HuP84}. In the electromagnetic field 
we perform a homodyne detection experiment. Its integrated 
photocurrent is the measurement result for measurement of 
the field-quadrature $A_t+A_t^*$ continuously in time. 
Using the characteristic functions introduced by Barchielli 
and Lupieri \cite{BaL85}, we find the probability density 
for these measurement results. In this density the $x$- and 
$z$-component of the Bloch vector of the initial state appear, 
indicating that homodyne detection is in fact a joint measurement 
of $\sigma_x$ and $\sigma_z$ in the initial state. 

The goal of the paper is to construct random variables (pointers) 
on the probability space of homodyne measurement results 
having distributions as close as possible  
to those of the ob\-ser\-va\-bles $\sigma_x$ and $\sigma_z$ in 
the initial state of the qubit.
`As close as possible' is taken to mean that the pointer must give an
unbiased estimate of the observable, with its worst case added variance
as low as possible.
Using an argument due to Wiseman \cite{Wis96},
we first show that optimal random variables will only depend 
on the endpoint of a weighted path of the integrated 
photocurrent. Allowed to restrict our attention to this smaller class of
pointers, we are able to use standard variational calculus to obtain the 
optimal random variables. They do not achieve the bound 
imposed by the Heisenberg-like relation for the worst case 
added variances \cite{Jan05}, but will be off by less than 5.6\%. 

The remainder of the paper is organized as follows. In 
Section \ref{sec model} we introduce the model of the 
qubit coupled to the field in the weak coupling limit. 
Section \ref{sec quality} introduces the quality of 
a measurement in terms of the worst case added variance. 
This section also contains the Heisenberg-like relation 
for joint measurement. In Section \ref{sec characteristic function}
we calculate the characteristic function of Barchielli 
and Lupieri for the homodyne detection experiment. 
Section \ref{sec variation} deals with the variational 
calculus to find the optimal pointers. In Section 
\ref{sec densities} we calculate the densities of the 
optimal pointers and then capture our main results 
graphically. 
In the last section we discuss our 
results.

\section{The model}\label{sec model}

We consider a two-level atom, i.e. a qubit, in interaction with 
the quantized electromagnetic field. The qubit is described 
by $\BB{C}^2$ and the electromagnetic field by the \emph{symmetric 
Fock space} $\FC$ over the Hilbert space of quadratically integrable 
functions $L^2(\BB{R})$ (space of one-photon wave functions), i.e.\
  \begin{equation*}
  \FC := \BB{C} \oplus \bigoplus_{k = 1}^\infty L^2(\BB{R})^{\ten_s k}.
  \end{equation*}
With the Fock space $\FC$ we can describe superpositions of field-states 
with different numbers of photons. The joint 
system of qubit and field together is described by the 
Hilbert space $\BB{C}^2\ten\FC$.

The interaction between the qubit and the electromagnetic field is 
studied in the weak coupling limit \cite{Gou04}, \cite{Gou05}, \cite{AFLu90}. 
This means that in the interaction picture the unitary dynamics of 
the qubit and the field together is given by a quantum 
stochastic differential equation (QSDE) in the sense of Hudson 
and Parthasarathy \cite{HuP84} 
  \begin{equation}\label{eq QSDE}
  dU_t =\Big\{\sigma_{-} dA_t^* - \sigma_{+} dA_t - 
  \half\sigma_{+} \sigma_{-} dt\Big\}U_t,
  \ \ \ \mbox{with}\ \ \ 
  \sigma_{-}  = \begin{pmatrix}0 & 0 \\ 1 & 0 \end{pmatrix}, \ 
  \sigma_{+}  = \begin{pmatrix}0 & 1 \\ 0 & 0 \end{pmatrix}, \
  U_0 = I. 
  \end{equation}    
The operators $\sigma_{-}$ and $\sigma_{+}$ are the annihilator and
creator on the two-level system. 
The field annihilation and creation processes
are denoted $A_t$ and $A^*_t$, respectively. 
Keep in mind that the evolution
$U_{t}$ acts nontrivially on the combined system $\BB{C}^2\ten\FC$, 
whereas $\sigma_{\pm}$ and $A_{t}$ are understood to designate
the single system-operators $\sigma_{\pm}\ten I$ and $I\ten A_{t}$.
Throughout
the paper we will remain in the interaction picture. Equation \eqref{eq QSDE} 
should be understood as a shorthand for the integral equation
  \begin{equation*}
  U_t = I + \int_0^t\sigma_{-} U_{\tau} dA^*_{\tau} 
   - \int_0^t\sigma_{+} U_{\tau}dA_{\tau} 
   - \half\int_0^t\sigma_{+} \sigma_{-} U_{\tau}d\tau, 
  \end{equation*} 
where the integrals on the right-hand side are 
stochastic integrals in the sense of Hudson and Parthasarathy \cite{HuP84}. 
The value of these integrals does not lie in 
their actual definition (on which we will not comment further), but in 
the It\^o rule satisfied by them, allowing for easy calculations.

\begin{stel}\label{Itorule}\textbf{(Quantum It\^o rule \cite{HuP84}, \cite{Par92})}
Let $X_t$ and $Y_t$ be stochastic integrals of the form 
\begin{equation*}\begin{split}
&dX_t = C_t dA_t + D_t dA_t^* + E_tdt \\
&dY_t = F_t dA_t + G_t dA_t^* + H_tdt
\end{split}\end{equation*}
for some stochastically integrable processes $C_t, D_t, E_t, F_t, 
G_t$ and $H_t$ (see \cite{HuP84}, \cite{Par92} for definitions). 
Then the process $X_tY_t$ satisfies the relation
  \begin{equation*}
  d(X_tY_t) = X_tdY_t + (dX_t)Y_t + dX_tdY_t,
  \end{equation*}
where $dX_tdY_t$ should be evaluated according to 
the quantum It\^o table:
\begin{center}
{\large \begin{tabular} {l|lll}
 & $dA_t$  & $dA^*_t$ & $dt$\\
\hline 
$dA_t$ & $0$ & $dt$ & $0$\\
$dA^*_t$ & $0$  & $0$ & $0$\\
$dt$ & $0$ & $0$ & $0$
\end{tabular} }
\end{center}
i.e.\ $dX_tdY_t = C_tG_tdt$. 
\end{stel}
As a corollary we have that, for any $f \in C^{2}(\BB{R})$, the 
process $f(X_{t})$ satisfies
$\textstyle d (f(X_{t})) = f'(X_{t})dX_{t} + \half f'' (X_{t}) (dX_{t})^2,$
where $(dX_t)^2$ should be evaluated according to the quantum It\^o table.

First a matter of notation. The quantum It\^o rule will be used for 
calculating differentials of products of stochastic integrals. 
Let $\{Z_i\}_{i=1,\dots, p}$ be stochastic integrals. Then we write 
 \begin{equation*} 
 d (Z_1Z_2\dots Z_p)= \sum_{\substack{\nu\subset\{1,\dots, p\} \\ \nu \neq \emptyset}}[\nu]
 \end{equation*}
where the sum runs over all {\it non-empty} subsets of $\{1,\dots, p\}$. For any 
$\nu=\{i_1,\dots, i_k\}$, the term $[\nu]$ is the contribution 
to $d (Z_1Z_2\dots Z_p)$ coming from differentiating only the terms with indices 
in the set $\{i_1,\dots, i_k\}$ and preserving the order of the factors in the product. 
The differential $d(Z_1Z_2Z_3)$, for example, contains terms of type
$[1]$, $[2]$, $[3]$, $[12]$, $[13]$, $[23]$ and $[123]$. We have
$[2]=Z_1(dZ_2)Z_3$, $[13]=(dZ_1)Z_2(dZ_3)$, $[123]=(dZ_1)(dZ_2)(dZ_3)$,
etc.

Let us return to equation \eqref{eq QSDE}. In order to illustrate how the 
quantum It\^o rule will be used,
we calculate the time evolution on the qubit explicitly. 
The algebra of qubit-observables is the algebra of 
$2 \times 2$-matrices, denoted $M_2(\BB{C})$.
The algebra of observables in the field is given by $B(\FC)$, the bounded operators
on $\FC$.
If $\mbox{id}:\ 
M_2(\BB{C})\to M_2(\BB{C})$ is the identity map and $\phi :\ B(\FC)\to\BB{C}$ is 
the expectation with respect to the vacuum state 
$\Phi := 1 \oplus 0 \oplus 0 \oplus \ldots \in \FC$ 
(i.e.\ $\phi (Y) := \langle \Phi, Y\Phi  \rangle$),
then time evolution on the qubit $T_t :\ M_2(\BB{C}) \to M_2(\BB{C})$ 
is given by
$ T_t(X) := \mbox{id}\ten\phi(U_t^*X\ten IU_t)$.
On the combined system, the full time evolution 
$j_t :\ M_2(\BB{C})\ten B(\FC) \to M_2(\BB{C})\ten B(\FC)$  
is given by $j_t(W) := U^*_tWU_t$.
In a diagram this 
reads 
   \begin{equation}\label{dildiag}\begin{CD}
  M_2 @>T_t>> M_2              \\
   @V{  \mbox{ \footnotesize id} \ten I}VV        @AA{\mbox{\footnotesize id} \ten \phi}A      \\
   M_2\ten B(\FC) @>j_t>> M_2\ten B(\FC).            \\
  \end{CD}\end{equation}
In the Schr\"odinger picture the arrows would be reversed.  
A qubit-state $\rho$ would be extended with the vacuum 
to $\rho\ten\phi$, then time evolved with $U_t$, and in the last 
step the partial trace over the field would be taken,
resulting in the state $\rho \circ T_t$.

Using the It\^o rule we can derive a (matrix-valued) differential 
equation for $T_t(X)$, i.e.\
  \begin{equation}\label{eq T}\begin{split}
  dT_t(X) &= \mbox{id}\ten\phi\big(d(U_t^*X\ten IU_t)\big) \\
  &= \mbox{id}\ten\phi\big((dU_t^*)X\ten IU_t + U_t^*X\ten I(dU_t) + 
  (dU_t^*)X\ten I(dU_t)\big) \\ 
  &= \mbox{id}\ten\phi\big(U_t^*L(X)\ten IU_t\big)dt \\
  &= T_t\big(L(X)\big)dt,
  \end{split}\end{equation}
where $L$ is the Lindblad generator 
  \begin{equation*}
  L(X) := -\half (\sigma_{+} \sigma_{-} X +X\sigma_{+} \sigma_{-} ) + \sigma_{+} X \sigma_{-} . 
  \end{equation*} 
In the derivation \eqref{eq T} we used the QSDE for $U^*_t$ which easily 
follows from \eqref{eq QSDE}
  \begin{equation*}
  dU^*_t = U_t^*\Big\{\sigma_{+} dA_t -\sigma_{-} dA^*_t - \half\sigma_{+} \sigma_{-} dt\Big\}, \ \ \ \ U^*_0 = I.
  \end{equation*} 
Furthermore, we used that stochastic integrals with respect to $dA_t$ and 
$dA^*_t$ vanish with respect to the vacuum expectation, leaving us only with the 
$dt$ terms. The differential equation \eqref{eq T} with initial 
condition $T_0(X) = X$ is solved by 
$T_t(X) =  \exp(tL)(X)$, which is exactly the 
time evolution of a two-level system spontaneously decaying to 
the ground state, as it should be. Although the arguments above 
are completely standard (cf.\ \cite{HuP84}), they do 
illustrate nicely and briefly some of the techniques used also in following 
sections.

\section{Quality of information transfer}\label{sec quality}

Now suppose we do a homodyne detection experiment, enabling 
us to measure the observables $A_t + A_t^*$ in the field 
continuously in time \cite{Bar90}.
If initially the qubit is in state 
$\rho$, then at time $t$ the qubit and field together are in  
a state $\rho^t$ on $M_2 (\BB{C}) \ten B(\FC)$ given by 
$\rho^t(W) := \rho\ten\phi(U^*_tWU_t) = 
\rho\big(\mbox{id}\ten\phi(U^*_tW U_t)\big)$.
Since 
  \begin{equation*}\begin{split}
  d\Big(\mbox{id}\ten\phi(U^*_t I\ten (A_{t} + A^*_{t}) U_t)\Big) &= 
  \mbox{id}\ten\phi\Big(d(U^*_t I\ten (A_{t} + A^*_{t}) U_t)\Big) \\
  &=\mbox{id}\ten\phi\Big([1]+[2]+[3]+[12]+[13]+[23]+[123]\Big) \\&=  
  \mbox{id}\ten\phi\Big(U_t^*(\sigma_{-}  + \sigma_{+} )\ten IU_t\Big)dt 
  \\&= 
  \exp(t L)(\sigma_{-} +\sigma_{+} )dt 
  = e^{-\frac{t}{2}}\sigma_xdt, 
  \end{split}\end{equation*}
we have that regardless the initial state $\rho$ of the qubit,  
the expectation of $(A_{t} + A^*_{t})$ in the final state $\rho^t$ will equal the 
expectation of $( 2 - 2e^{-\frac{t}{2}} )\sigma_x$ in the initial state $\rho$.

\subsection{Defining the quality of information transfer}

The process at hand is thus a transfer of information about $\sigma_x$ 
to a `pointer' $A_t + A^*_t$, which can be read off by means of homodyne
detection. This motivates the following definition.

\begin{de}\textbf{(Unbiased Measurement \cite{Jan05})}
Let X be an observable of the qubit, i.e.\ a self-adjoint 
element of $M_2(\BB{C})$, and let $Y$ be an observable of 
the field, i.e.\ a self-adjoint operator in (or affiliated to) $B(\FC)$. 
An \emph{unbiased measurement} 
$M$ of $X$ with \emph{pointer} $Y$ is by definition a completely 
positive map $M:\ B(\FC) \to M_2(\BB{C})$ such that $M(Y) = X$. 
\end{de}  

Needless to say, for each fixed $t$ the map $M:\ B(\FC) \to M_2(\BB{C})$ given by 
$M(B) := \mbox{id}\ten\phi(U^*_t I\ten BU_t)$  is a measurement of $\sigma_x$ with pointer 
$Y = (2-2e^{-\frac{t}{2}})^{-1}(A_t + A_t^*)$. This means that, 
after the measurement procedure of coupling to the field in the 
vacuum state and allowing for interaction with the qubit for $t$ 
time units, the distribution of the measurement 
results of the pointer $Y$ has inherited the expectation of 
$\sigma_x$, regardless of the initial state $\rho$. However, 
we are more ambitious and would like its distribution as a whole 
to resemble that of $\sigma_x$. This motivates the following 
definition. 

\begin{de}\textbf{(Quality \cite{Jan05})} 
Let $M:\ B(\FC) \to M_2(\BB{C})$ be an unbiased measurement of 
$X$ with pointer $Y$. Then its \emph{quality} $\sigma$ is defined by
  \begin{equation*}
  \sigma^2 := \mbox{sup}\Big\{\mbox{Var}_{\rho\circ \hspace{-0.2 mm} M}(Y) - \mbox{Var}_\rho(X)\big|\ 
  \rho \in \mathcal{S}(M_2)\Big\},
  \end{equation*}
where $\mathcal{S}(M_2)$ denotes the state space of $M_2(\BB{C})$ (i.e.\ all 
positive normalized linear functionals \mbox{on $M_2(\BB{C})$}).
\end{de}

This means that $\sigma^2$ is the variance added to the initial 
distribution of $X$ by the measurement procedure $M$ for the worst 
case initial state $\rho$. A small calculation shows that
  \begin{equation*}
  \mbox{Var}_{\rho\circ \hspace{-0.2 mm} M}(Y) - 
  \mbox{Var}_\rho(X) = \rho\big(M(Y^2)-M(Y)^2\big), 
  \end{equation*}
which implies that $\sigma^2 = \| M(Y^2) -  M(Y)^2 \| $, where 
$X \mapsto \| X \|$
denotes the operator norm \mbox{on $M_2(\BB{C})$}. In particular this shows 
that $\sigma^2$ is positive, as one might expect. It follows from \cite{Jan05} that
$\sigma$ equals zero if and only if the measurement procedure $M$ exactly carries over 
the distribution of $X$ to $Y$. In short, $\sigma$ is a suitable measure for 
how well $M$ transfers information about $X$ to the pointer $Y$. 

\subsection{Calculating the quality of information transfer}

Let us return to the example at hand, i.e.\ 
$M(B) = \mbox{id}\ten\phi(U_t^*I\ten B U_t)$, with field-observable $Y = (2-2e^{-\frac{t}{2}})^{-1}(A_t + A_t^*)$ 
as a pointer for $\sigma_x$. Let us calculate its quality, which amounts to
evaluating $M(Y^2) = 
(2-2e^{-\frac{t}{2}})^{-2}M\big((A_t+A_t^*)^2\big)$. 
To this aim,
we will first introduce some ideas which will be 
of use to us in later calculations as well. 

\begin{de}\label{def F} 
Let $f$ and $h$ be real valued functions, $h$ twice 
differentiable. Let $Y_t$ be given by $dY_t = f(t)(dA_t + dA_t^*)$, 
$Y_0 = 0$. For $X \in M_2(\BB{C})$ we define 
 \begin{equation*}
 F_h(X,t) := \mbox{id}\ten\phi\big(U_t^* X \ten h(Y_t)U_t\big). 
 \end{equation*}
When no confusion can arise we shall shorten $F_h(X,t)$ to $F_h(X)$.
\end{de}

The homodyne detection experiment has given us a measurement 
result (the integrated photocurrent) which is just the path of 
measurement results for $A_t + A_t^*$ continuously in time. Given 
this result, we post-process it by weighting the increments of the 
path with the function $f(t)$ and letting 
$h(y)$ act on the result. 
The following lemma will considerably shorten calculations.

\begin{lem}\label{lem eq F}
  \begin{equation*}
  \frac{dF_h(X)}{dt} = F_h\big(L(X)\big) + f(t)F_{h'}(\sigma_{+} X +
  X\sigma_{-} ) + \half f(t)^2F_{h''}(X)
  \end{equation*}
\end{lem} 
\begin{proof}
Using the notation below Theorem \ref{Itorule} with $Z_1 = U_t^*$, 
$Z_2 = I \ten h(Y_t)$ and $Z_3 = U_t$, we find
  \begin{equation*}
  dF_h(X) = \mbox{id}\ten\phi\big([1] + [2] + [3] + [12] + [13] + [23] + [123]\big). 
  \end{equation*}
Again we will use that the vacuum expectation kills all $dA_t$ and $dA^*_t$ terms.   
Using Theorem \ref{Itorule} we see that after the vacuum expectation 
the terms $[1]$, $[3]$ and $[13]$ make up $F_h\big(L(X)\big)dt$. Since 
third powers of increments are $0$ we again have $[123] = 0$. From 
  \begin{equation*}
  dh(Y_t) = h'(Y_t)f(t)(dA_t + dA^*_t) + \half h''(Y_t)f(t)^2 dt,
  \end{equation*}
we find that, after taking vacuum expectations, the terms $[12]$ and $[23]$ make up
the second term $f(t)F_{h'}(\sigma_{+} X + X\sigma_{-} )dt$ and $[2]$ provides the 
last term $\frac{1}{2}f(t)^2F_{h''}(X)dt$.      
\end{proof}

We are now well-equipped to calculate $M\big((A_t+A_t^*)^2\big)$.
Choose $f(t) = 1$ and $h(x) = x^2$. (The maps $x \mapsto x^n$ will be denoted $\x^n$
hereafter.) Then
$M\big((A_t+A_t^*)^2\big) = F_{\x^2}(I)$ and by Lemma \ref{lem eq F}
  \begin{equation}\label{eq Fx2}
  \frac{dF_{\x^2}(I)}{dt} = 2 F_{\x}(\sigma_{-} +\sigma_{+} ) + F_{1}(I) = 
  2 F_{\x}(\sigma_{-} +\sigma_{+} ) + I, \ \ \ \ 
  F_{\x^2}(I, 0) = 0. 
  \end{equation}
Applying Lemma \ref{lem eq F} to 
$F_{\x}(\sigma_{-} +\sigma_{+} )$, we obtain  
  \begin{equation}\label{eq Fx}
  \frac{dF_{\x}(\sigma_{-} +\sigma_{+} )}{dt} = -\half F_{\x}(\sigma_{-} + \sigma_{+} ) + 
  2F_{1}(\sigma_{+} \sigma_{-} ), \ \ \ \
  F_{\x}(\sigma_{-} +\sigma_{+} , 0) = 0.
  \end{equation}
Finally, $F_{1}(\sigma_{+} \sigma_{-} )$ satisfies
  \begin{equation} \label{eq Fx0}
  \frac{dF_{1}(\sigma_{+} \sigma_{-} )}{dt} = - F_{1}(\sigma_{+} \sigma_{-} ),\ \ \ \ 
  F_{1}(\sigma_{+} \sigma_{-} , 0) = \sigma_{+} \sigma_{-}.   
  \end{equation}
Solving \eqref{eq Fx0}, \eqref{eq Fx} and \eqref{eq Fx2} successively leads
first to  
$F_1(\sigma_{+} \sigma_{-} ) = e^{-t}\sigma_{+} \sigma_{-} $, then to
$F_{\x}(\sigma_{-} +\sigma_{+} ) = 4(e^{-\frac{t}{2}} - e^{-t})\sigma_{+} \sigma_{-} $
and finally to
$F_{\x^2}(I) = 8(e^{-\frac{t}{2}} -1)^2 \sigma_{+} \sigma_{-}  + tI$. 
Consequently, the quality of the measurement $M$ of $\sigma_x$ 
with pointer $Y = (2-2e^{-\frac{t}{2}})^{-1}(A_t+ A_t^*)$ 
is given by
  \begin{equation*} \begin{split}
  \sigma^2 & = \| M(Y^2)- M(Y)^2 \| = 
  \left\| \frac{8(e^{-\frac{t}{2}} -1)^2 \sigma_{+} \sigma_{-}  + tI}
  {(2-2e^{-\frac{t}{2}})^{2}} - I \right\| = 
  \left\| 2\sigma_{+} \sigma_{-}  + 
  \left(\frac{t}{(2-2e^{-\frac{t}{2}})^{2}} - 1\right)I \right\| \\
  & = \frac{t}{(2-2e^{-\frac{t}{2}})^{2}} + 1.
  \end{split}\end{equation*}
This expression takes its minimal value $2.228$ at $t = 2.513$, 
leading to a quality $\sigma = 1.493$. 

The calculation above has an interesting side product.
The observable $M\big( (A_t + A_t^*)^2 \big)$ depends linearly on 
$\sigma_{z}$, indicating that in addition to information
on $\sigma_{x}$, also information on $\sigma_z$ in the initial
qubit-state ends up in the measurement outcome. 
Indeed, if we use as a pointer
  \begin{equation}\label{eq tilde Y}
  \tilde{Y} := \frac{(A_t+A_t^*)^2 -tI}{4 (e^{-\frac{t}{2}}-1)^2} - I,  
  \end{equation}
then we have $M(\tilde{Y}) = \sigma_z$, so that 
$M$ is also 
a measurement of $\sigma_z$ with pointer $\tilde{Y}$.   
  
Note that the pointers $Y$ and $\tilde{Y}$ commute, i.e.\ measuring 
$A_t+A_t^*$ via the homodyne detection scheme is an indirect 
\emph{joint measurement} of $\sigma_x$ and $\sigma_z$. If we would 
also like to gain some information about $\sigma_y$, we could 
for example sweep the measured quadrature through $[0, 2\pi)$ in time 
by measuring $e^{i\omega t}A_t + e^{-i\omega t}A^*_t$ instead. In this paper however, 
we will restrict ourselves to continuous time measurement of $A_t + A_t^*$,
as additional information on $\sigma_y$ would deteriorate the quality of 
$\sigma_x$- and/or $\sigma_z$-measurement.  
The following theorem is a Heisenberg-like relation that 
gives a bound on how well joint measurements can be performed.

\begin{stel}\textbf{(Joint Measurement \cite{Jan05})}\label{thm joint}
Let $M:\ B(\FC) \to M_2(\BB{C})$ be an unbiased measurement 
of self-adjoint observables $X \in M_2(\BB{C})$ and $\tilde{X} \in M_2(\BB{C})$ 
with self-adjoint commuting pointers 
$Y$ and $\tilde{Y}$ in (or affiliated to) $B(\FC)$, respectively. Then for 
their corresponding qualities $\sigma$ and $\tilde{\sigma}$ 
the following relation holds
  \begin{equation*}
  2\sigma\tilde{\sigma} \ge \| [X, \tilde{X}]\|.
  \end{equation*}
\end{stel}

Denote by $\tilde{\sigma}$ the quality of the $\sigma_z$
measurement with the pointer $\tilde{Y}$ defined in \eqref{eq tilde Y}.
Since $[\sigma_x,\sigma_z] = -2i\sigma_y$, the qualities $\sigma$ and 
$\tilde{\sigma}$ (corresponding to the pointers $Y$ and $\tilde{Y}$, respectively)
satisfy the inequality 
  \begin{equation}\label{eq ineq} 
  \sigma\tilde{\sigma} \ge 1. 
  \end{equation}
Using similar techniques as before, that is recursively calculating 
$F_{\x^4}(I)$ via Lemma \ref{lem eq F}, we find 
  \begin{equation*}
  \tilde{\sigma}^2 = \frac{t^2}{8(e^{-\frac{t}{2}}-1)^4} + 
  \frac{2t-4(e^{-\frac{t}{2}}-1)^2}{(e^{-\frac{t}{2}}-1)^2}.
  \end{equation*}
This expression takes its minimal value $8.836$ at $t = 2.513$. This 
leads to a quality $\tilde{\sigma} = 2.973$, which means that 
$\sigma\tilde{\sigma} = 4.437$, i.e.\ we are far removed from hitting the 
bound $1$ in \eqref{eq ineq}. However, there is still some room for 
manoeuvring by post-processing 
of the homodyne measurement data.

\section{The weighted path}\label{sec characteristic function}

Let us presently return to our homodyne detection experiment. 
We observe $A_{\tau} +A_{\tau}^*$ continuously in time, i.e.\ the result of our 
measurement is a path $\omega$ of measurement results $\omega_{\tau}$ (the 
photocurrent integrated up to time $\tau$) for $A_{\tau}+A_{\tau}^*$.
This means that we have a space $\Omega$ of all possible measurement 
paths and that we can identify an operator $A_{\tau}+A_{\tau}^*$ with the 
map from $\Omega$ to $\BB{R}$ mapping a measurement 
path $\omega \in \Omega$ to the measurement result $\omega_{\tau}$ at time $\tau$.     
That is, we have simultaneously diagonalized the family of commuting 
operators $\{A_{\tau}+A^*_{\tau}|\ \tau \ge 0\}$ and viewed them as random variables on 
the spectrum $\Omega$. The spectral projectors of the operators 
$\{A_{\tau}+A_{\tau}^*|\ 0\le \tau \le t\}$ endow $\Omega$ with a filtration 
of $\sigma$-algebras $\Sigma_t$. Furthermore, the 
states $\rho^{\tau}$, defined by $\rho^{\tau} (W) := \rho\ten\phi(U^*_{\tau} W U_{\tau})$ 
provide a family of consistent measures $\BB{P}_{\tau}$ on $(\Omega,\Sigma_{\tau})$, 
turning it into the probability space $(\Omega,\Sigma_t,\BB{P})$. 
(See e.g. \cite{BT4}.)

We aim to find random variables on $(\Omega,\Sigma_t,\BB{P})$ 
having distributions resembling those of $\sigma_x$ and $\sigma_z$ in 
the initial state $\rho$. In the previous section we used the 
random variables 
  \begin{equation} \label{naive choice}
  Y(\omega) = \frac{\omega_{\tau}}{2-2e^{-\frac{\tau}{2}}} \ \ \ \ \mbox{and} \ \ \ \
  \tilde{Y}(\omega) =\frac{\omega_{\tau}^2 -{\tau}}{4 (e^{-\frac{\tau}{2}}-1)^2} - 1,
  \ \ \ \ \tau = 2.513
  \end{equation}
for $\sigma_x$ and $\sigma_z$, respectively. Our next goal is to 
find the optimal  
random variables, in the sense of the previously defined quality. 


\subsection{Restricting the class of pointers}

In our specific example, $M$ is given by $M(B) = 
\mbox{id}\ten\phi(U_{\tau}^* I\ten B U_{\tau})$. Note that stochastic 
integrals with respect to the annihilator $A_{\tau}$ acting 
on the vacuum vector $\Phi$ are zero. Therefore, we can modify 
$U_{\tau}$ to $Z_{\tau}$, given by
  \begin{equation*}
  dZ_{\tau} = \Big\{\sigma_{-} (dA^*_{\tau}+dA_{\tau}) - 
  \half \sigma_{+} \sigma_{-} d\tau \Big\}Z_{\tau},\ \ \ \  Z_0 = I, 
  \end{equation*}
without affecting $M$ \cite{Bel92a}.  
Therefore, for all $B \in B(\FC)$, we have $M(B) = \mbox{id}\ten\phi(U^*_{\tau} I\ten
BU_{\tau}) =$ 
\mbox{$\mbox{id}\ten\phi(Z^*_{\tau} I\ten B Z_{\tau})$}. The solution 
$Z_t$ can readily be found, it is given by
  \begin{equation*}
  Z_t = \begin{pmatrix}e^{-\frac{1}{2}t} & 0 \\ 
  \int_0^te^{-\frac{1}{2}\tau}(dA_{\tau}+dA_{\tau}^*) & 1 \end{pmatrix}.
  \end{equation*}
Note that $Z_t$, as a matrix valued function of the measurement 
path, is an element of $M_2(\BB{C}) \ten \CC_t$, where $\CC_t$ is 
the commutative von Neumann algebra generated by 
$A_{\tau} + A_{\tau}^*,\ 0\le \tau \le t$.
Moreover we see that $Z_t$ is not a function of all the $(A_{\tau}+A_{\tau}^*)$'s 
separately, it is only a function of the endpoint of the weighted path 
$Y_{t} = \int_0^t e^{-\frac{1}{2}\tau} (dA_{\tau}+dA_{\tau}^*)$ \cite{Wis96}.
Therefore if we define $\mathcal{S}_t \subset \CC_t$ to be 
the commutative von Neumann algebra generated by $Y_t$,
then we even have $Z_{t} \in M_2(\BB{C}) \ten \mathcal{S}_t$. 

Denote by $C \mapsto \BB{E}[\, C \,|\SC_t]$ the unique classical 
conditional expectation from $\CC_t$ onto $\SC_t$ that 
leaves $\phi$ invariant, i.e.\ $\phi(\BB{E}[\,C\,|\SC_t]) = \phi(C)$ for 
all $C \in \CC_t$.
We can extend $ \BB{E}[\ \cdot \ |\SC_t]$ by tensoring it 
with the identity map on the $2 \times 2$ matrices 
to obtain a map $\mbox{id}\ten \BB{E}[\ \cdot \ |\SC_t]$ from 
$M_2(\BB{C}) \ten \CC_t$ onto $M_2(\BB{C})\ten \SC_t$. 
From the positivity of $\BB{E}[\ \cdot \ |\SC_t]$ as 
a map between commutative algebras, it follows that 
$\mbox{id} \ten \BB{E}[\ \cdot \ |\SC_t]$ is completely positive.
Since $\BB{E}[\ \cdot \ |\SC_t]$ satisfies $\BB{E}[\,CS\,|\SC_t] = 
\BB{E}[\,C\,|\SC_t] S$ for all $C \in \CC_t$ and $S \in \SC_t$, 
we find that $\mbox{id} \ten \BB{E}[\ \cdot \ |\SC_t]$ satisfies 
the \emph{module property}, i.e. 
  \begin{equation*}
  \mbox{id} \ten \BB{E}[\ \cdot \ |\SC_t](A_1BA_2) 
  = A_1\Big(\mbox{id}\ten\BB{E}[\ \cdot \ |\SC_t](B)\Big)A_2, 
  \end{equation*}
for all $A_1, A_2 \in M_2(\BB{C})\ten\SC_t$ and 
$B \in M_2(\BB{C})\ten\CC_t$. Moreover, if $\rho$ is a state on 
$M_2(\BB{C})$, then it follows from the invariance of $\phi$ under 
$\BB{E}[\ \cdot \ |\SC_t]$ that $\mbox{id}\ten\BB{E}[\ \cdot \ |\SC_t]$ 
leaves $\rho \ten \phi$ invariant. We conclude that, given $\rho$ 
on $M_2(\BB{C})$, the map 
$\mbox{id}\ten\BB{E}[\ \cdot \ |\SC_t]$ from $M_2(\BB{C})\ten\CC_t$ onto 
$M_2(\BB{C})\ten \SC_t$ is the 
unique conditional expectation in the noncommutative 
sense of \cite{Tak71} that leaves $\rho\ten\phi$ 
invariant. We will use the shorthand $\BB{E}_{\SC_t}$ for 
$\mbox{id}\ten \BB{E}[\ \cdot \ |\SC_t]$ in the following.

\begin{lem} \label{essence}
Let $C \in \CC_t $ be a pointer with quality $\sigma_{C}$ such that $M(C) = X$. Then 
$\tilde{C} := \BB{E}[\,C\, | \mathcal{S}_t]$ is 
also a pointer with $M(\tilde{C}) = X$,
and with quality $\sigma_{\tilde{C}} \le \sigma_{C}$.
\end{lem}
\begin{proof}
Note that for all states $\rho$ on $M_2(\BB{C})$ we have
  \begin{equation*}\begin{split}
  \rho\big(M(\tilde{C})\big) &= \rho\ten\phi\big(Z_t^*I \ten \tilde{C} Z_t\big) = 
  \rho\ten\phi\Big(Z_t^* \BB{E}_{\SC_t} (I\ten C) Z_t\Big) 
  =\rho\ten\phi\Big(\BB{E}_{\SC_t} \big(Z_t^* I\ten C Z_t\big)\Big) 
  \\&=\rho\ten\phi(Z_t^* I \ten C Z_t) = \rho\big(M(C)\big) = \rho(X),
  \end{split}\end{equation*}
where we used the module property and the fact that $Z_t$ is an 
element of $M_2\ten \SC_t$ in the third step and the invariance 
of $\rho \ten \phi$ in the fourth step. Since this holds for 
all states $\rho$ on $M_2(\BB{C})$, we conclude
that $M(\tilde{C}) = X$. 

As for the variance, we note first that the conditional expectation 
$\BB{E}_{\SC_t}$ is a completely positive identity preserving map.
Therefore,
for all self-adjoint $C \in \CC_t$, we have
  \begin{equation*}
  \BB{E}_{\SC_t}(I\ten C^2) \ge 
  \Big(\BB{E}_{\SC_t}(I\ten C)\Big)^2
  \end{equation*}
by the Cauchy-Schwarz inequality for completely positive maps \cite{Wer01}.

We can now apply the same strategy as before. For all states $\rho$ on 
$M_2(\BB{C})$ we have 
  \begin{eqnarray*}
  \rho\big(M(C^2)\big) &=& \rho \ten \phi (Z_t^* I \ten C^2 Z_t) \\
  &=& 
  \rho \ten \phi \Big(\BB{E}_{\SC_t}
  \big(Z_t^* I \ten C^2  Z_t\big)\Big)\\
  &=&  
  \rho \ten \phi \Big(Z_t^*\BB{E}_{\SC_t}(I \ten C^2) Z_{t}\Big)\\
  &\geq& 
  \rho \ten \phi \Big( Z_t^*\Big(\BB{E}_{\SC_t}(I \ten C)\Big)^2Z_{t}\Big)\\ 
  &=& \rho\big(M(\tilde{C}^2)\big) \,.
  \end{eqnarray*}
Thus $M(C^2) \geq M(\tilde{C}^2)$, and in particular
$\sigma_{C}^2 = \| M(C^2) - M(C)^2 \| \geq \| M(\tilde{C}^2) - M(\tilde{C})^2 \|
= \sigma_{\tilde{C}}^2$.
\end{proof}

This has a very useful consequence: 
if we are looking for pointers that record, say 
$\sigma_{x}$ or $\sigma_{z}$ in an optimal fashion, 
then it suffices to examine 
only pointers in $\mathcal{S}_t$.
Instead of sifting through the collection of all random variables 
on the measurement outcomes, we are thus allowed to 
confine the scope of our search to the 
rather transparent collection of  
measurable functions of $Y_{t}$. 
In the following, we will look at such pointers $h_t(Y_{t})$. 
We will usually drop the subscript $t$ on $h$ to 
make the notation lighter. 

\subsection{Distribution of $Y_{t}$}

At this point we are interested in the 
probability distribution of the random variable $Y_t$. 
Its characteristic function \cite{BaL85} is given by
  \begin{equation*}
  E(k) := \BB{E}_{\rho^t}\big[\exp(-ik Y_t)\big] = 
  \rho \ten \phi \Big(U_t^* I \ten \exp(-ik Y_t)U_t\Big) = 
  \rho \left(F_{\exp(-ik \x)}(I)\right), 
  \end{equation*}  
so that we need only calculate $F_{\exp(-ik \x)}(I)$. For notational 
convenience we will replace the subscript $\exp(-ik \x)$ by $k$ in the
following. Using Lemma \ref{lem eq F}, we find the following system of
matrix valued differential equations:
  
  \begin{tabular}{l c l}
  $\displaystyle \frac{dF_k(I)}{dt} $ \rule[-4,2 mm]{0 mm}{5 mm} &=&$
  \displaystyle -ik e^{-\frac{t}{2}}F_k(\sigma_{-} +\sigma_{+} ) -
  \frac{k^2e^{-t}}{2} F_k(I),$\\ 
  $\displaystyle \frac{dF_k(\sigma_{-} +\sigma_{+} )}{dt} $ \rule[-4,2 mm]{0 mm}{5 mm} 
  &=&$ \displaystyle -\half F_k(\sigma_{-} +\sigma_{+} ) 
  - 2ike^{-\frac{t}{2}} F_k(\sigma_{+} \sigma_{-} ) 
  - \frac{k^2 e^{-t}}{2}F_k(\sigma_{-}  + \sigma_{+} ), $\\
  $\displaystyle \frac{dF_k(\sigma_{+} \sigma_{-} )}{dt} $ \rule[-4,2 mm]{0 mm}{5 mm}&=&$
  \displaystyle -F_k(\sigma_{+} \sigma_{-} ) - 
  \frac{k^2e^{-t}}{2}F_k(\sigma_{+} \sigma_{-} ),$\\
  with initially& & $\displaystyle F_k(I,0) = I,\ \ \ \
  F_k(\sigma_{-} +\sigma_{+} , 0) = \sigma_{-} +\sigma_{+} ,\ \ \ \
   F_k(\sigma_{+} \sigma_{-} , 0) = \sigma_{+} \sigma_{-} .$
  \end{tabular}

Solving this system leads to
  \begin{equation*}
  F_k(I) = e^{-\frac{k^2(1-e^{-t})}{2}}\Bigg(I  
  -ik \big(1-e^{-t}\big)\big(\sigma_{-} + \sigma_{+} \big) - 
  k^2\big(1-e^{-t}\big)^2\sigma_{+} \sigma_{-} \Bigg).
  \end{equation*} 
We define the Fourier transform to be $\FC(f)(x) := 
\frac{1}{\sqrt{2\pi}}\int_{-\infty}^\infty f(k) e^{ikx}dk$. Then 
the probability density of $Y_t$ with respect to the Lebesgue
measure is given by $\frac{1}{\sqrt{2\pi}} \FC(E)(x) = 
\frac{1}{\sqrt{2\pi}} \rho\Big(\FC\big(F_k(I)\big)(x)\Big)$. 
Defining $p(x):=\frac{1}{\sqrt{2\pi}} \FC\big(F_k(I)\big)(x)$, 
we can write
  \begin{equation*}
  p(x) =  \frac{e^{-\frac{1}{2}\frac{x^2}{1-e^{-t}}}}{\sqrt{2\pi(1-e^{-t})}}
  \Big(I + x(\sigma_{-} +\sigma_{+} ) +
   (x^2-1+e^{-t})\sigma_{+} \sigma_{-} \Big),
  \end{equation*}
i.e.\ $Y_t$ is distributed according to a Gaussian perturbed by the matrix 
elements of the initial state $\rho(\sigma_{-} +\sigma_{+} ) =  \rho(\sigma_x)$ and $\rho(\sigma_{+} \sigma_{-} ) = 
\frac{1}{2}\rho(\sigma_z)+\frac{1}{2}$. No information about $\rho$ on 
$\sigma_y$ enters the distribution though. To gain information about $\sigma_y$ 
we would have to change our continuous time measurement setup, 
as we discussed before. If we absorb a constant $(1-e^{-t})^{-\frac{1}{2}}$ in 
the definition of $Y_t$, i.e.\ $Y_t := 
(1-e^{-t})^{-\frac{1}{2}}\int_0^t
e^{-\frac{\tau}{2}}(dA_{\tau}+dA_{\tau}^*)$,  
then its density becomes
  \begin{equation}\label{eq p}
  p(y) =  \frac{e^{-\frac{y^2}{2}}}{\sqrt{2\pi}}
  \Big(I + \beta_ty(\sigma_{-} +\sigma_{+} ) +
   \beta_t^2(y^2-1)\sigma_{+} \sigma_{-} \Big),
  \end{equation}
where $\beta_t : = \sqrt{1-e^{-t}}$.

\section{Variational calculus} \label{sec variation}

In Lemma \ref{essence}, we have shown that it suffices to 
consider only random variables of the form $h(Y_t)$ for some
measurable $h$.
In equation \eqref{eq p}, we have captured the probability 
distribution of $Y_t$.
All that remains now is to 
calculate the optimal $h$, which can be done with variational 
calculus. 

\subsection{Optimal $\sigma_x$-measurement}

We seek the function $h^*$ for which 
the quality $\sigma$ of the pointer $h^*(Y_t)$ for 
$\sigma_x$-measurement is optimal.
 In other words, we need  
\begin{equation}\label{eq sigma2}
  \sigma^2 := \left\| \int_{-\infty}^\infty h^2(y)p(y)dy - 
  \Big(\int_{-\infty}^\infty h(y)p(y)dy\Big)^2\right\|
  := \left\| \begin{pmatrix} d_1 & 0 \\ 0 & d_2 \end{pmatrix}  \right\|
  \end{equation}  
to be minimal under the restriction 
$\int_{-\infty}^\infty h(y)p(y)dy = \sigma_{x}$.

Now $\sigma^2$ is the norm of a diagonal $2 \times 2$-matrix with entries $d_1$ and $d_2$.
Both depend smoothly on $h$, but $\sigma^2 = \mbox{max}\{ d_1 , d_2 \}$ does not.
There are three possibilities: 
\begin{itemize}
\item[{\footnotesize I)}] $\sigma^2 = d_1$ in some open neighborhood of $h^*$. 
	   To find these $h^*$, we must minimize the smooth functional $d_1$ and then
	   check whether $d_1 < d_2$. 	  
\item[{\footnotesize II)}] $\sigma^2 = d_2$ in some open neighborhood of $h^*$. To find these $h^*$, we must 
	   minimize $d_2$ and check whether $d_2 < d_1$.  
\item[{\footnotesize III)}] $d_1 = d_2$ for $h^*$. To find these $h^*$, we must minimize $d_1$ subject to the condition
         $d_1 = d_2$. 
\end{itemize} 
In principle, we need three different functionals $\Lambda_1$, $\Lambda_2$ and $\Lambda_3$ 
for these three distinct cases. 
However, it turns out that we can make due with the following functional
 \begin{equation}\label{eq tildeLambda}\begin{split} 
  \Lambda(h,\kappa,\gamma_1,\gamma_2,\gamma_3) :=\ 
  &\Big( \frac{1}{\sqrt{2\pi}}\int_{-\infty}^\infty h^2(y)e^{-\frac{1}{2}y^2}dy - 1\Big) + 
  \kappa\Big(\frac{\beta_t^2}{\sqrt{2\pi}}
  \int_{-\infty}^\infty h^2(y)(y^2-1)e^{-\frac{1}{2}y^2}dy\Big)\ + \\
  &\gamma_1\Big(\int_{-\infty}^\infty h(y)e^{-\frac{1}{2}y^2}dy\Big) +   
  \gamma_2\Big(\int_{-\infty}^\infty h(y)(y^2-1)e^{-\frac{1}{2}y^2}dy\Big)\ + \\
  &\gamma_3\Big(\frac{\beta_t}{\sqrt{2\pi}}
  \int_{-\infty}^\infty h(y)ye^{-\frac{1}{2}y^2}dy - 1\Big).
  \end{split}\end{equation} 
The constants 
$\gamma_1, \gamma_2$ and $\gamma_3$ are the Lagrange multipliers enforcing 
$\int_{-\infty}^\infty h(y)p(y)dy = \sigma_{x}$. 
These are needed in all cases: $\Lambda_1$, $\Lambda_2$ and $\Lambda_3$.
One can readily check that setting $\kappa = 0$ in $\Lambda$ yields $\Lambda_1$,
setting $\kappa = 1$ yields $\Lambda_2$ 
and considering $\kappa$ as a free Lagrange
multiplier forces $d_1 = d_2$, so that one has $\Lambda_3 = \Lambda$. 

All three cases lead to similar optimality conditions.
The requirement that the optimal solution be stable under first order variations 
yields $h^*$ satisfying either
\begin{equation}\label{one}
  h^*(x) = \frac{C_1 x + C_2}{x^2 + \varepsilon} + C_3 
\end{equation}
or
\begin{equation}\label{two}
  h^*(x) = C_{4}x^2 + C_{5}x + C_{6} \,     
\end{equation}
for some real constants $C_1,C_2,C_3,C_4, C_5, C_6$ and $\varepsilon$ depending 
on $\kappa,\gamma_1,\gamma_2,\gamma_3$.

Suppose that $h^*$ takes the form (\ref{two}). The constraint
$\int_{-\infty}^{\infty} h^*(y) p(y) dy = \sigma_{-}  + \sigma_{+} $ will then force
$C_{4} = C_{6} = 0$ and $C_{5} = \beta_{t}^{-1}$, so that 
$h^*(y) = \beta_{t}^{-1} y$. The random variable 
we are investigating is simply the observed path, weighted by the function 
$f(\tau) = \beta_t^{-1}e^{-\tau /2}$, with $t$ the final time of measurement.
Since all the integrals we encounter are Gaussian moments, we can readily compute 
$M(h^{*2}(Y_t)) = \int_{-\infty}^{\infty} h^{*2}(y) p(y) dy \ $ 
to be $ \ 2\sigma_{+} \sigma_{-}  + \beta_{t}^{-2} I$.
Thus $$\sigma^2 = \|(2\sigma_{+} \sigma_{-}  + \beta_{t}^{-2} I ) 
- (\sigma_{-} +\sigma_{+} )^2  \| = 1 + \beta_{t}^{-2}\,.$$
For $t \to \infty$, this amounts to $\sigma \to \sqrt{2}$.
Already, we have improved on the naive result $\sigma = 1.493$ 
obtained previously.

We proceed with the more involved case (\ref{one}), which will 
provide us with the optimal solution.
Before we continue with the constants 
$C_1,C_2,C_3$ and $\varepsilon$ however,
we calculate some integrals for later use.

\begin{de}  
Define the error function $\mbox{erf}(x)$ and integrals $I(\varepsilon)$ and $J(\varepsilon)$ by
  \begin{equation*}
  \mbox{erf}(x) := \frac{2}{\sqrt{\pi}}\int_{0}^x e^{- u^2}du,\ \ \ \
  I(\varepsilon) := \int_{-\infty}^\infty \frac{e^{-\frac{x^2}{2}}}{x^2+\varepsilon}dx,
  \ \ \ \
  J(\varepsilon) := \int_{-\infty}^\infty \frac{e^{-\frac{x^2}{2}}}{(x^2+\varepsilon)^2}dx.
  \end{equation*}
\end{de}

\begin{lem}\label{integrals}
  \begin{equation*}
  J(\varepsilon) = \frac{\sqrt{2\pi} + (1-\varepsilon)I(\varepsilon)}{2\varepsilon} \ \ \ \
  \mbox{and} \ \ \ \ 
  I(\varepsilon)= \pi \sqrt{\frac{e^{\varepsilon}}{\varepsilon}}
  \left( 1  - \mbox{erf}\left(\sqrt{\frac{\varepsilon}{ 2}}\right) \right).
  \end{equation*}
\end{lem}
\begin{proof}
Since the Fourier transform of $e^{-\sqrt{\varepsilon}|k|}$ is equal to 
$\sqrt{\frac{2\varepsilon}{\pi}}\frac{1}{x^2 + \varepsilon}$, we find
  \begin{equation*}\begin{split}
  I(\varepsilon) &= \sqrt{\frac{\pi}{2\varepsilon}}\int_{-\infty}^\infty 
  \FC\big(e^{-\sqrt{\varepsilon}|k|}\big)\FC\big(e^{-\frac{k^2}{2}}\big)dx = 
  \sqrt{\frac{\pi}{2\varepsilon}}\int_{-\infty}^\infty 
  e^{-\sqrt{\varepsilon}|k|}e^{-\frac{k^2}{2}}dk\ 
  = \sqrt{\frac{2\pi}{\varepsilon}}\int_{0}^\infty 
  e^{-\sqrt{\varepsilon}k}e^{-\frac{k^2}{2}}dk \\&=
  \sqrt{\frac{2\pi}{\varepsilon}}e^{\frac{1}{2}\varepsilon}\int_{\sqrt{\varepsilon}}^\infty
  e^{-\frac{u^2}{2}}du = \pi \sqrt{\frac{e^{\varepsilon}}{\varepsilon}}
  \left( 1 - \mbox{erf}\left(\sqrt{\frac{\varepsilon}{2}}\right)\right),
  \end{split}\end{equation*}
where, in the second step, we have used that the Fourier transform $\FC$ is unitary.
The expression for $J$ follows from 
  \begin{equation*}\begin{split}
  0 & = \frac{xe^{-\frac{x^2}{2}}}{x^2+\varepsilon}\Bigg|_{-\infty}^\infty = 
  \int_{-\infty}^\infty
  \frac{d}{dx}\left(\frac{xe^{-\frac{x^2}{2}}}{x^2+\varepsilon}\right)dx =
  \int_{-\infty}^\infty 
  \left(\frac{1-x^2}{x^2+\varepsilon} -
  \frac{2x^2}{(x^2+\varepsilon)^2}\right)e^{-\frac{x^2}{2}}dx  \\
  & = \int_{-\infty}^\infty \left(-1 + \frac{\varepsilon -1}{x^2 + \varepsilon} + 
  \frac{2\varepsilon}{(x^2+\varepsilon)^2}\right)e^{-\frac{x^2}{2}}dx = 
  -\sqrt{2\pi} + (\varepsilon -1)I(\varepsilon) + 2\varepsilon J(\varepsilon).
  \end{split}\end{equation*}
\end{proof}
The condition $\int_{-\infty}^\infty h^*(y)p(y)dy = \sigma_{-}  + \sigma_{+}  =\sigma_x$ implies
  \begin{equation*}
  C_1 = \frac{\sqrt{2\pi}}{\beta_t\big(\sqrt{2\pi}- \varepsilon I(\varepsilon)\big)}, 
  \ \ \ \ C_2 = C_3 = 0
  \end{equation*}
which fixes $C_1$ as a function of $\varepsilon$. 
The next step is to express $d_1$ and $d_2$ in terms of $\varepsilon$:
  \begin{equation*}\begin{split}
  d_2 &= \frac{C_1^2}{\sqrt{2\pi}}\int_{-\infty}^
  \infty \frac{y^2}{(y^2+\varepsilon)^2}e^{-\frac{y^2}{2}}dy - 1
  = \frac{C_1^2}{\sqrt{2\pi}}\Big(I(\varepsilon)-\varepsilon J(\varepsilon)\Big)-1,\\
  d_1 & = \frac{C_1^2\beta_t^2}{\sqrt{2\pi}}\int_{-\infty}^\infty
  \frac{y^2(y^2-1)}{(y^2 + \varepsilon)^2}e^{-\frac{y^2}{2}}dy + d_2= 
  \frac{C_1^2\beta_t^2}{\sqrt{2\pi}}\Big(\sqrt{2\pi} - (1+2\varepsilon)I(\varepsilon)+ 
  \varepsilon(1+\varepsilon)J(\varepsilon)\Big) + d_2.
  \end{split}\end{equation*}
First, we use Lemma \ref{integrals} to express the above in terms of elementary functions
and the error function. Then, using Maple, we find that $\varepsilon \mapsto \mbox{max}\{d_1,d_2\}$ 
has a unique minimum at $\varepsilon = 0.605$, 
for which $d_1 =d_2 = 0.470$.
This leads to 
a $C_1$ that equals 2.359,
and to a quality of 
  \begin{equation*}
  \sigma = \sqrt{\mbox{max}\{d_1,d_2\}} = 0.685 \,.
  \end{equation*}  

\subsection{Optimal $\sigma_z $-measurement}

For optimal $\sigma_z$-measurement, we can run the same program.  
We search for the function $\tilde{h}$ which optimizes the quality 
$\tilde\sigma$, under the restriction that $\tilde{h}(Y_t)$ 
be a pointer for $\sigma_z$-measurement. 
That is, we search for a function $\tilde{h}$ minimizing 
the functional of equation \eqref{eq sigma2}, but now 
under the restriction $\int_{\infty}^\infty h(y)p(y)dy = \sigma_z$.
Again there are three cases of interest, $d_1=d_2$, $d_1>d_2$ and $d_2>d_1$, 
which we can treat simultaneously by introducing, analogous to  
equation \eqref{eq tildeLambda}, the functional
  \begin{equation*}\begin{split} 
  \tilde{\Lambda}(h,\kappa,\gamma_1,\gamma_2,\gamma_3) :=\ & 
  \Big( \frac{1}{\sqrt{2\pi}}\int_{-\infty}^\infty h^2(y)e^{-\frac{1}{2}y^2}dy - 1\Big) + 
  \kappa\Big(\frac{\beta^2_t}{\sqrt{2\pi}}\int_{-\infty}^\infty h^2(y)(y^2-1)e^{-\frac{1}{2}y^2}dy\Big)\ + \\
  &\gamma_1\Big(\frac{1}{\sqrt{2\pi}}\int_{-\infty}^\infty h(y)
  e^{-\frac{1}{2}y^2}dy+1\Big) +   
  \gamma_2\Big(\frac{\beta^2_t}{\sqrt{2\pi}}\int_{-\infty}^\infty h(y)(y^2-1)e^{-\frac{1}{2}y^2}dy-2\Big)\ + \\
  &\gamma_3\Big(\int_{-\infty}^\infty h(y)ye^{-\frac{1}{2}y^2}dy\Big).
  \end{split}\end{equation*}
Indeed, $\gamma_1, \gamma_2$ and $\gamma_3$ are the Lagrange multipliers 
enforcing the restriction $\int_{\infty}^\infty h(y)p(y)dy = \sigma_z$. 
Again, the functional $\sigma^2$ of equation \eqref{eq sigma2} depends 
non-differentiably on $h$ when $d_1=d_2$. We then have to search for the 
optimum among the points of non-differentiability, in which case 
$\kappa$ is the Lagrange multiplier confining us to these points. If $d_1 >d_2$
then $\kappa = 1$ and if $d_2 > d_1$ then 
$\kappa = 0$. Summarizing, wherever $\Lambda$ takes its minimal value, 
optimality implies $\frac{\delta\tilde\Lambda}{\delta h}
(\tilde{h}, \kappa, \gamma_1, \gamma_2, \gamma_3) = 0$ for some 
$\kappa, \gamma_1, \gamma_2$ and $\gamma_3$. Performing the 
functional derivative 
yields either
  \begin{equation}\label{one1}
  \tilde{h}(x) = \frac{D_1x+D_2}{x^2 + \delta} + D_3  
  \end{equation}
or
\begin{equation}\label{two2}
  \tilde{h}(x) = D_{4} x^2 + D_{5}x + D_{6}  
\end{equation}   
for some (time-dependent) constants $D_1, D_2, D_3, D_4, D_5, D_6$ and 
$\delta$ depending on 
$\kappa, \gamma_1, \gamma_2$ and $\gamma_3$. 

Again, we begin with the least demanding case \eqref{two2}, resulting from $\kappa = 0$.
The condition  $\int_{-\infty}^\infty \tilde{h}(y)p(y)dy = \sigma_z$ implies 
$D_{5} = 0$, $D_{4} = \beta_{t}^{-2}$ and $D_{6} = -1 -\beta_{t}^2$.
For $t \to \infty$,
this leads to $$\sigma^2 = \| M(\tilde{h}^2(Y_{t})) - M(\tilde{h}(Y_{t}))^2 \|
= \|(4 \sigma_{+} \sigma_{-}  + 3 I) - I  \| = 6\,,$$ so that $\sigma \to \sqrt{6}$.

This improves the result $\tilde{\sigma} = 2.973$ obtained 
previously, but once again the ultimate bound will be 
reached in the more arduous case \eqref{one1}.
There, the condition 
$\int_{-\infty}^\infty \tilde{h}(y)p(y)dy = \sigma_z$ implies 
  \begin{equation*}
  D_1\big(\sqrt{2\pi}-\delta I(\delta)\big) = 0,\ \ \ \ 
  D_2 = \frac{2\sqrt{2\pi}}{\beta_t^2\big(\sqrt{2\pi}-(1+\delta)I(\delta)\big)},\ \ \ \ 
  D_3 = - \frac{\sqrt{2\pi} + I(\delta)D_2}{\sqrt{2\pi}}. 
  \end{equation*}
This leads to expressions for $d_1$ and $d_2$ as a function of $\delta$.
Using Lemma \ref{integrals} and Maple once more, we find that the function $\delta \mapsto \mbox{max}\{d_1,d_2\}$ 
has a unique minimum at $\delta = 2.701$,
for which $d_1=d_2 = 2.373$. This leads to a quality of
  \begin{equation*}
  \tilde\sigma = \sqrt{\mbox{max}\{d_1,d_2\}} = 1.540,
  \end{equation*}  
  attained for $D_1 = 0$, $D_2 = -21.649$ and $D_3 = 5.391$.
For the joint measurement this leads to
  \begin{equation*}
  \sigma\tilde\sigma = 1.056\,.
  \end{equation*}
Although we did not achieve the bound of 1 provided by Theorem \ref{thm joint},
we have come as close as the measurement setup allows.
We conclude that, 
using the setup investigated in this article, no simultaneous measurement 
of $\sigma_x$ and $\sigma_z$ will be able to approach the quantum bound 
by more than 5.6 \%.  
Furthermore, we have identified the unique pointers for this optimal 
measurement in equations \eqref{one} and \eqref{one1}.

\section{Distribution of pointer variables}\label{sec densities}

We have designed pointers $h^*(Y_t)$ and 
$\tilde{h}(Y_t)$ in such a way that their distributions in the final state
best resemble the distributions of $\sigma_x$
and $\sigma_z$ in the initial state.
We will now calculate and plot these final densities. 

\subsection{Calculation of $h^*$- and $\tilde{h}$-densities}

Let $\rho$ be the initial state of 
the qubit and let it be parameterized by its Bloch vector $(P_x,P_y,P_z)$. 
By equation \eqref{eq p}, the density $q(y)$ of $Y_t$ is given by
  \begin{equation}\label{eq q}
  q(y) = \rho\big(p(y)\big) = \frac{e^{-\frac{1}{2}y^2}}{\sqrt{2\pi}}
  \Big(1 + \beta_t yP_x + \beta_t^2(y^2-1)\frac{P_z+1}{2}\Big).
  \end{equation} 

We are interested in the the distributions $r(x)$ and $s(x)$ 
of $h^*(Y_t)$ and $\tilde{h}(Y_t)$ respectively. 
Let us start with $h^*$. From equation 
\eqref{one}, we first calculate the points $y$ 
where $h^*(y) = x$ for some fixed value of $x$.  
  \begin{equation*}
  y_{\pm} = \frac{C_{1}\pm\sqrt{C_{1}^2-4x^2\varepsilon}}{2x} 
  \end{equation*}
By the Frobenius-Peron equation (see e.g.\ \cite{Ott}), $r(x)$ is given by
  \begin{equation*}
  r(x) = \sum_{+,-} \frac{q(y_\pm)}{|{h^*}'(y_\pm)|}, 
  \end{equation*}
which leads immediately to
  \begin{equation} \label{rofx}
  r(x) = \sum_{+,-} \frac{(y_\pm^2+\varepsilon)^2
  \Big(1 + \beta_t y_\pm P_x + \beta_t^2(y_\pm^2-1)\frac{P_z+1}{2}\Big)}{C_{1}|y^2_\pm -\varepsilon|} 
  \frac{e^{-\frac{1}{2}y_\pm^2}}{\sqrt{2\pi}}, 
  \end{equation}
where it is understood that $r(x) \neq 0$ only for $x \in
[-\frac{C_{1}}{2 \sqrt{\varepsilon}}, \frac{C_{1}}{2 \sqrt{\varepsilon}}]$. 

We run a similar analysis for $s(x)$. The points $y$ in which 
$\tilde{h}(y)= x$ are given by 
  \begin{equation*}
  y_\pm = \pm \sqrt{\frac{(x-D_3)\delta-D_2}{D_3-x}}.
  \end{equation*}
This leads to 
  \begin{equation} \label{sofz}
  s(x) = \sum_{+,-} 
  \frac{(y_\pm^2+\delta)^2
  \Big(1 + \beta_t y_\pm P_x + \beta_t^2(y_\pm^2-1)\frac{P_z+1}{2}\Big)}{2|D_2y_\pm|}
  \frac{e^{-\frac{1}{2}y_\pm^2}}{\sqrt{2\pi}},  
  \end{equation}  
with $s(x) \neq 0$ only for $x \in [D_{3} + \frac{D_{2}}{\delta}, D_{3}]$. 
We proceed with a graphical illustration of the results obtained so far. 

\subsection{Plots of $\sigma_{x}$-measurement}

According to formula \ref{eq q}, the 
distribution of the endpoint of the weighted path depends on the input qubit-state. 
For instance, the negative $\sigma_{x}$-eigenstate, the tracial state and
the positive $\sigma_{x}$-eigenstate lead to the
distributions below: \\[-0.7cm]
\begin{center}
\begin{tabular}{l c r}
\begin{minipage}[t]{4.5cm}
\epsfig{file=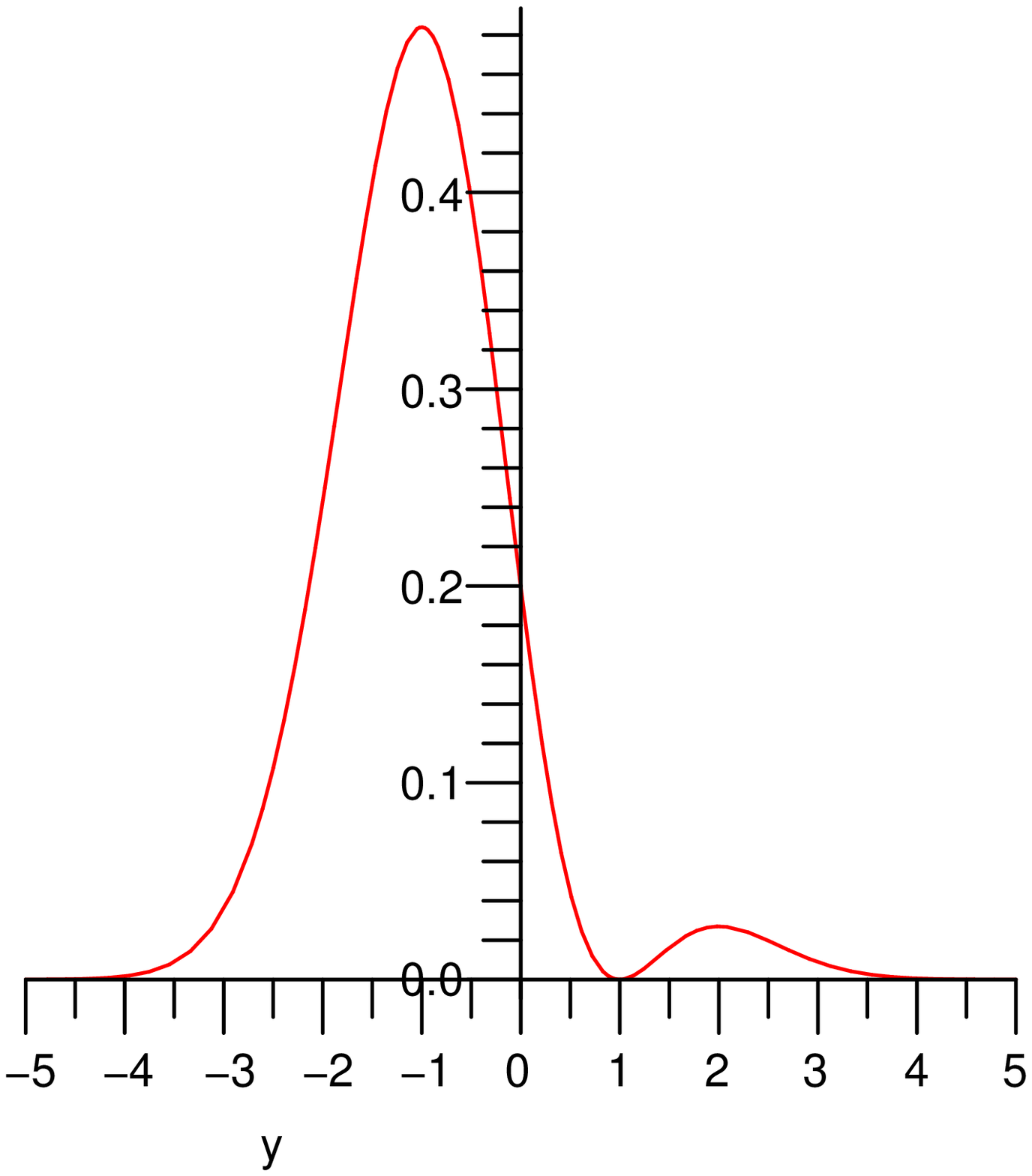, width=4.5cm}
\\[-2mm]{\footnotesize Figure 1: Probability density of the endpoint of the weighted path 
 for input $|\leftarrow \, \rangle $.}
\label{fig: g(y)-100}
\end{minipage}
&
\begin{minipage}[t]{4.5cm}
\epsfig{file=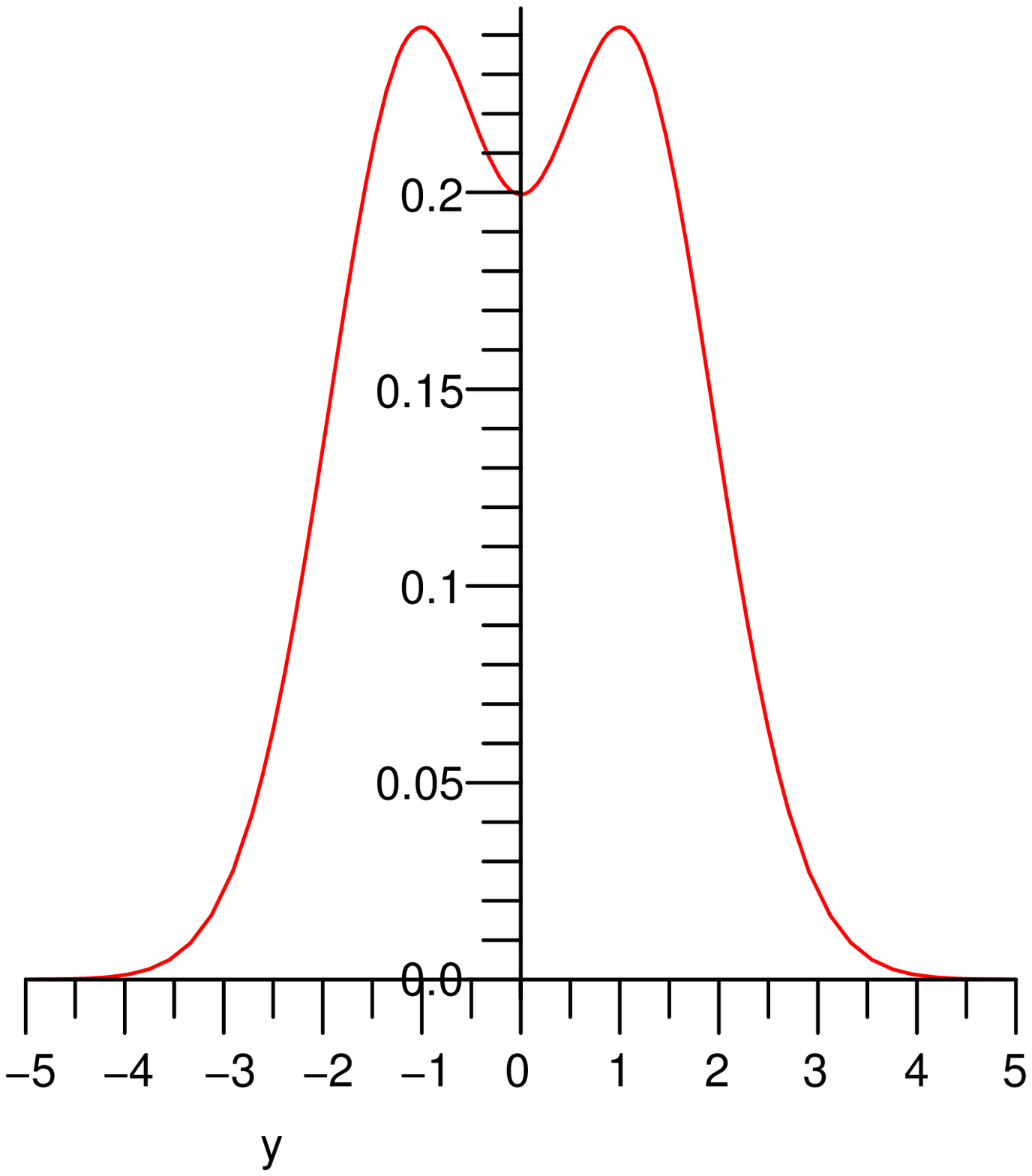, width=4.5cm}
\\[-2mm]{\footnotesize Figure 2: Probability density of the endpoint of the weighted path 
 for input $\mbox{tr}$.}
\label{fig: g(y)000}
\end{minipage}
&
\begin{minipage}[t]{4.5cm}
\epsfig{file=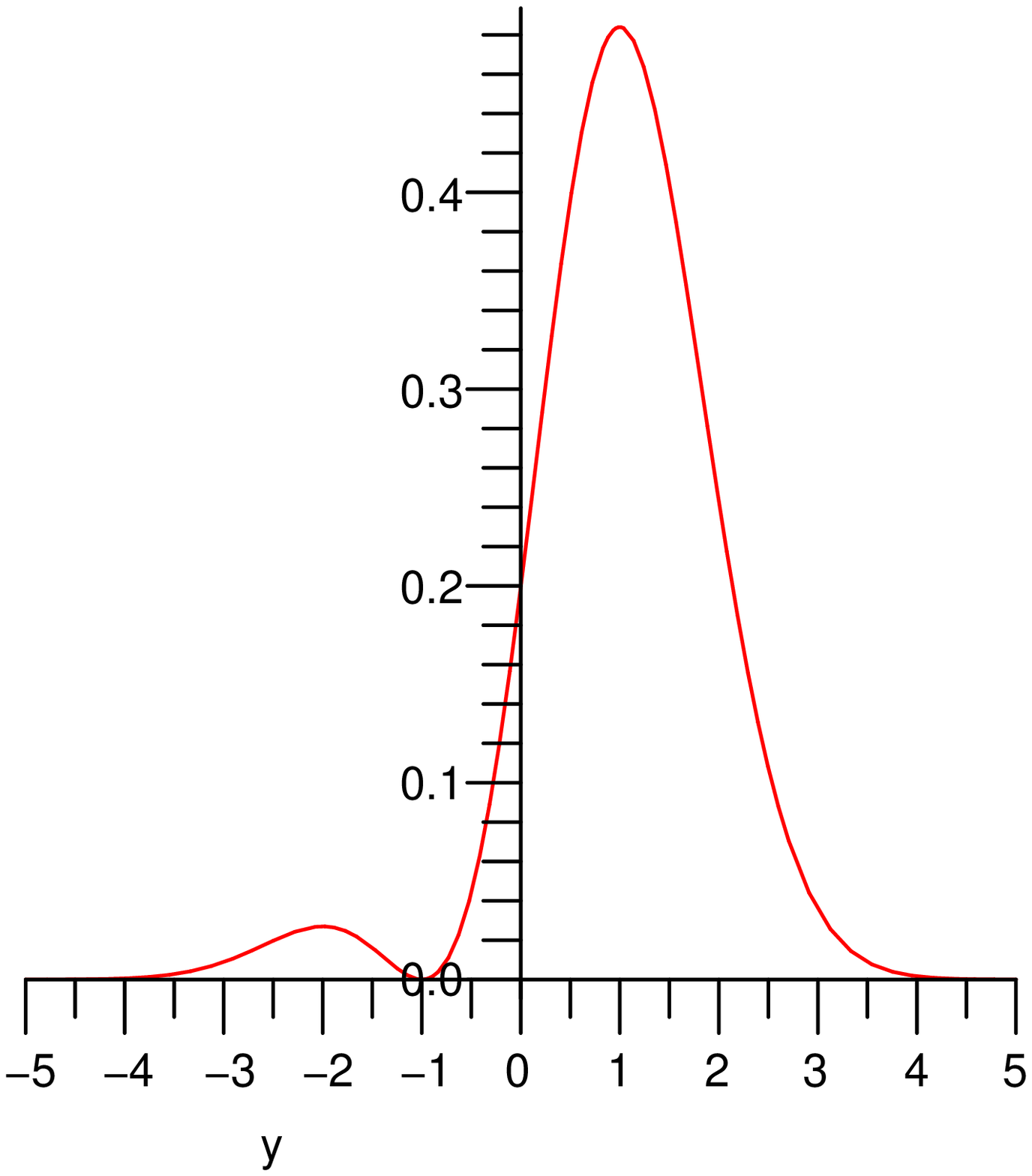, width=4.5cm}
\\[-2mm]{\footnotesize Figure 3: Probability density of the endpoint of the weighted path 
 for input $|\rightarrow \, \rangle $.}
\end{minipage}\\
\end{tabular}\\[-1cm]
\end{center}
%
%
%

\begin{tabular}{l c r}
\begin{minipage}[t]{9cm}
\mbox{}\\[3mm]
In order to estimate $\sigma_x$, we use the pointer
of $\sigma_x$ given by 
\begin{equation}\label{hstar}
h^*(x) = \frac{C_{1} x}{x^2 + \varepsilon},
\end{equation}
with
$\varepsilon = 0.605$ 
and
$C_{1} = 2.359$. It is illustrated in figure 4 to the right. 

In formula \eqref{rofx}, we have calculated the probability distributions of this
pointer under the distributions of the endpoint of the weighted path illustrated in figures
1, 2 and 3. They are illustrated in figures 5, 6 and 7 below.
\end{minipage}
&
\hspace{0.1cm}
&
\begin{minipage}[t]{4.5cm}
\mbox{}\\*
\epsfig{file=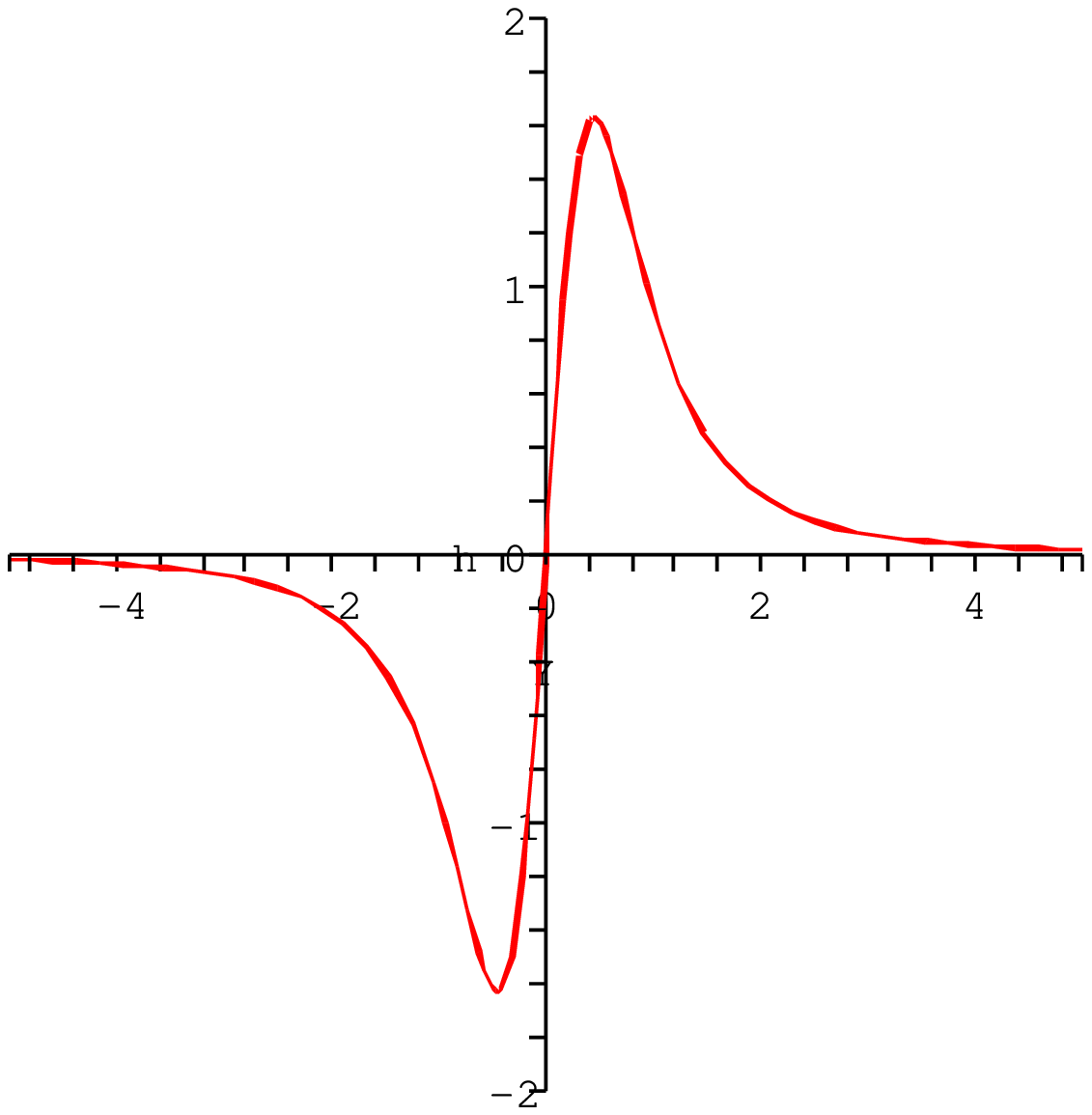, width=4.5cm}\label{fig: Pointer x}
 \\[-2mm] {\footnotesize Figure 4: Pointer for $\sigma_x$}\\
\mbox{}  
  \end{minipage}\\
\end{tabular}\\*
%
\begin{center}
\begin{tabular}{r c l}
 
\begin{minipage}[t]{4.5 cm}
  \epsfig{file=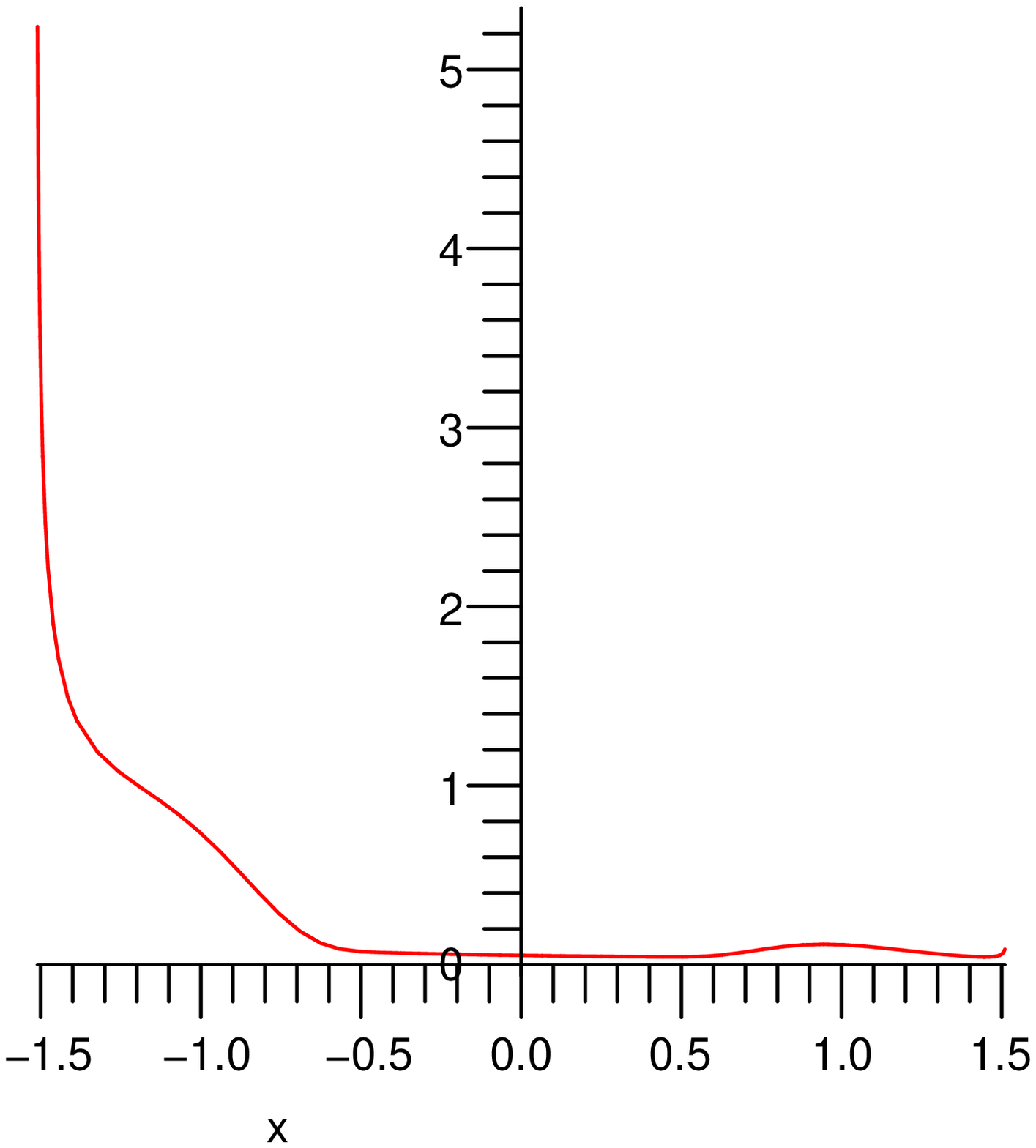, width=4.5cm}
  \\[-2mm] {\footnotesize Figure 5: Probability density of the $\sigma_x$-pointer 
  for input $|\leftarrow \, \rangle $.}
\end{minipage}
&
\begin{minipage}[t]{4.5 cm}
  \epsfig{file=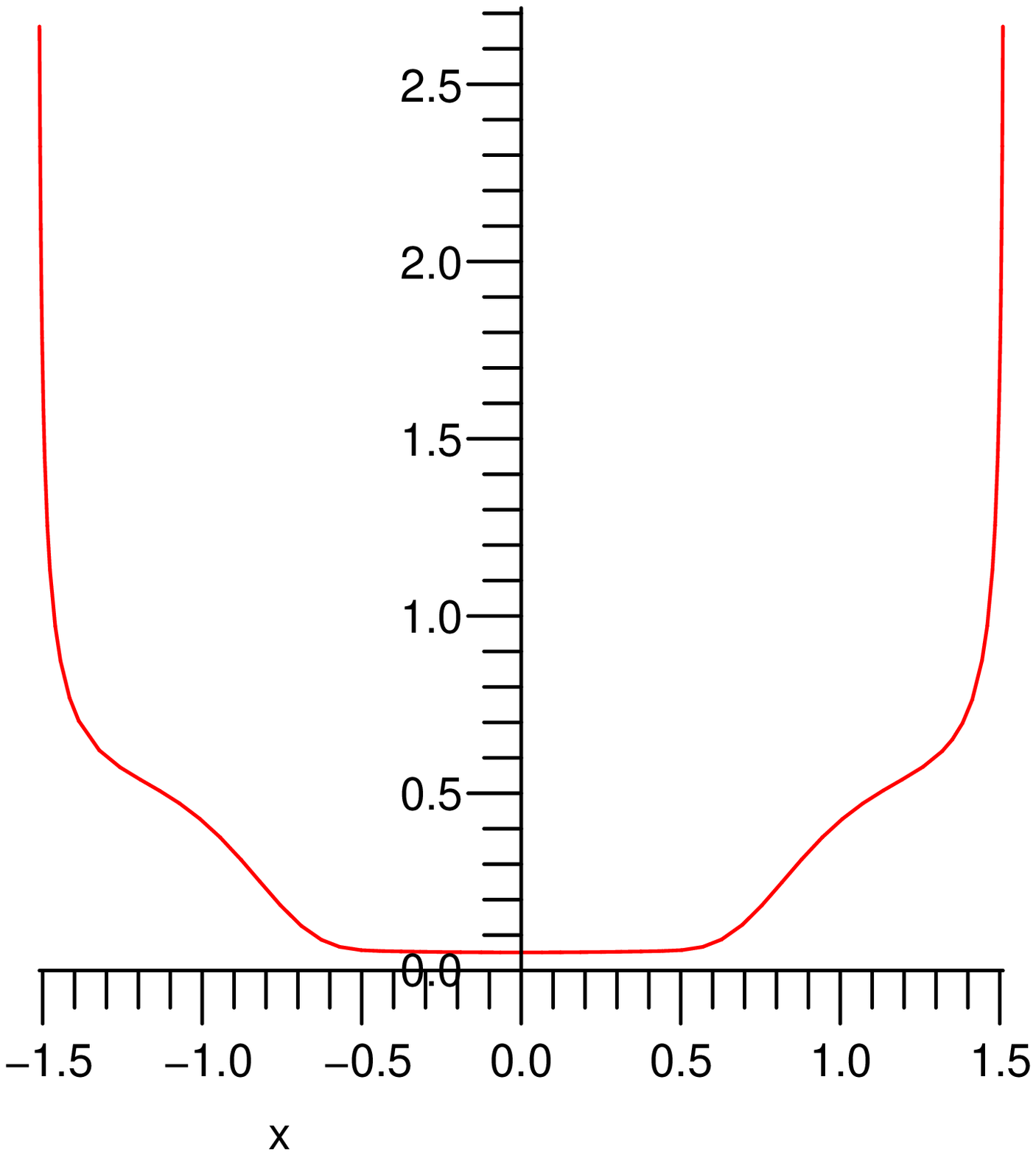, width=4.5cm}
 \\[-2mm] {\footnotesize Figure 6: Probability density of the $\sigma_x$-pointer 
  for input $\mbox{tr}$.}
\end{minipage}
&
\begin{minipage}[t]{4.5 cm}
  \epsfig{file=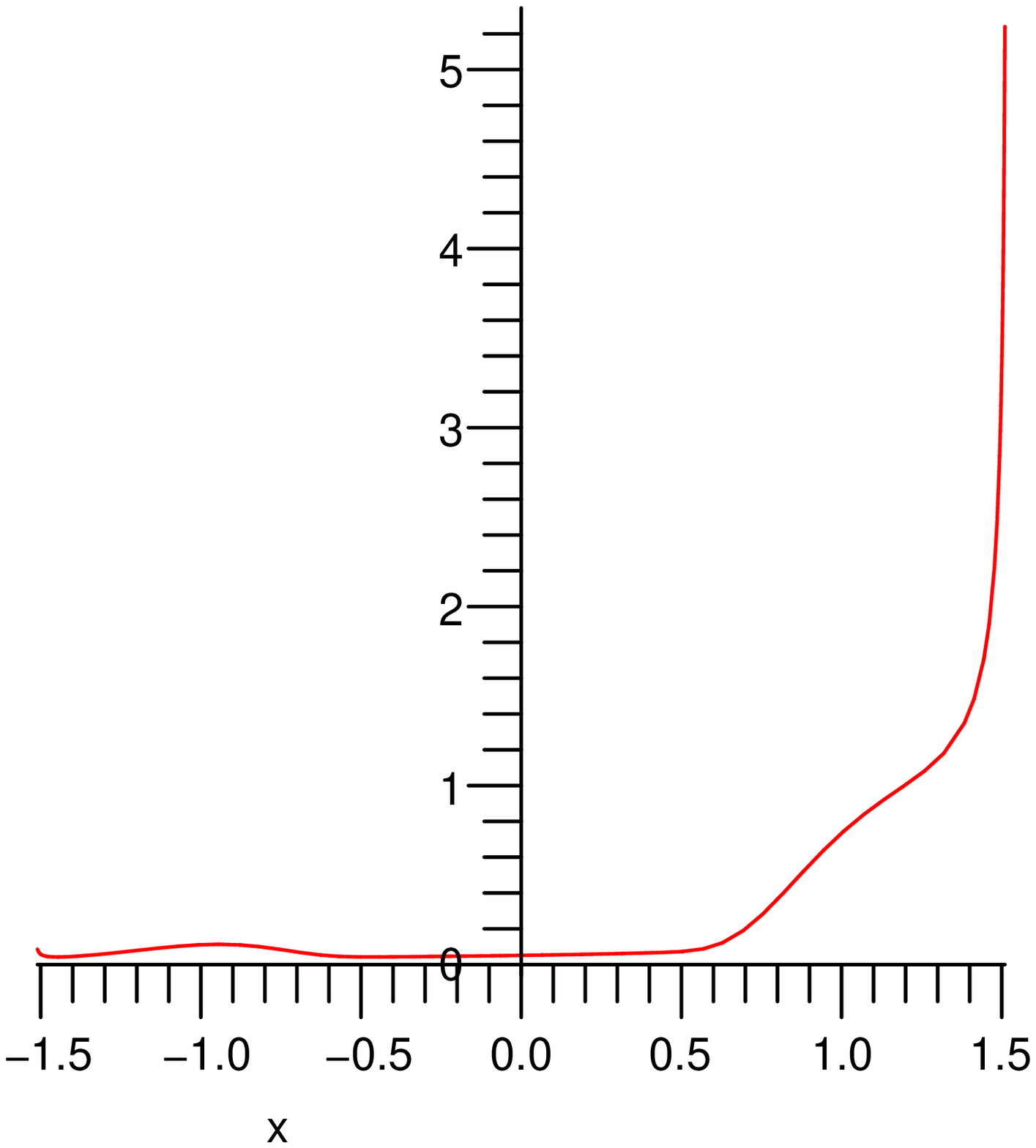, width=4.5cm}
 \\[-2mm] {\footnotesize Figure 7: Probability density of the $\sigma_x$-pointer 
  for input $|\rightarrow \, \rangle $.}
  \label{fig: r(x)100}
\end{minipage}\\

\end{tabular}
\end{center}

\subsection{Plots of $\sigma_{z}$-measurement} \label{plotz}

We repeat this for the $\sigma_{z}$-pointer.  
By formula \ref{eq q}, the positive $\sigma_{z}$-eigenstate, the
tracial state and the negative $\sigma_{z}$-eigenstate 
lead to the distributions of the endpoint of the weighted path shown below:\\[-0.7cm] 
\begin{center}
\begin{tabular}{l c r}
\begin{minipage}[t]{4.5cm}
  \epsfig{file=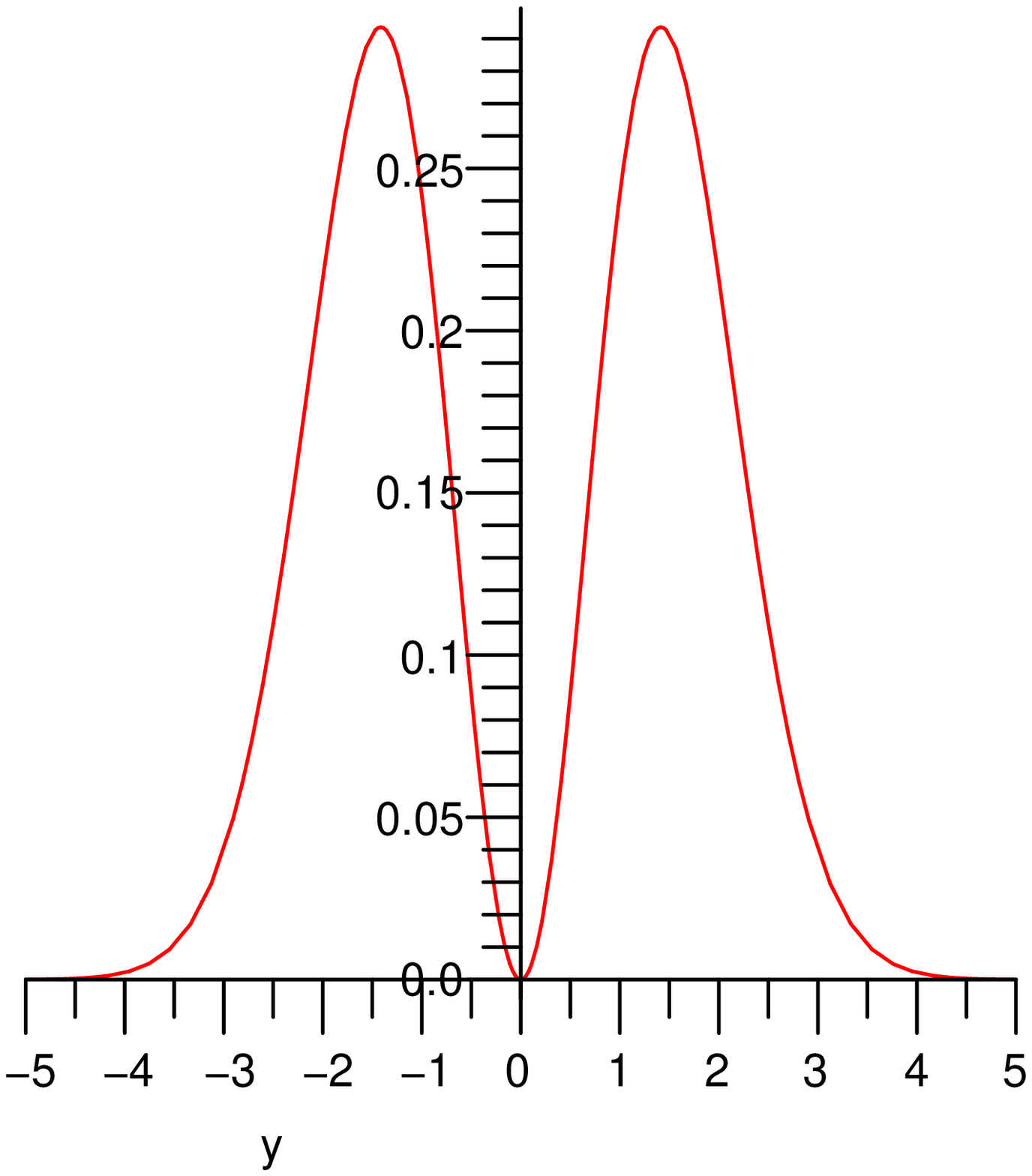, width=4.5cm}
 \\[-2mm] {\footnotesize Figure 8: Probability density of the endpoint of the weighted path 
  for input $|\uparrow \, \rangle $.}
\end{minipage}
&
\begin{minipage}[t]{4.5cm}
  \epsfig{file=qy000, width=4.5cm}
\\[-2mm]  {\footnotesize Figure 9: Probability density of the endpoint of the weighted path 
  for input $\mbox{tr}$.}
\end{minipage}
&
\begin{minipage}[t]{4.5cm}
  \epsfig{file=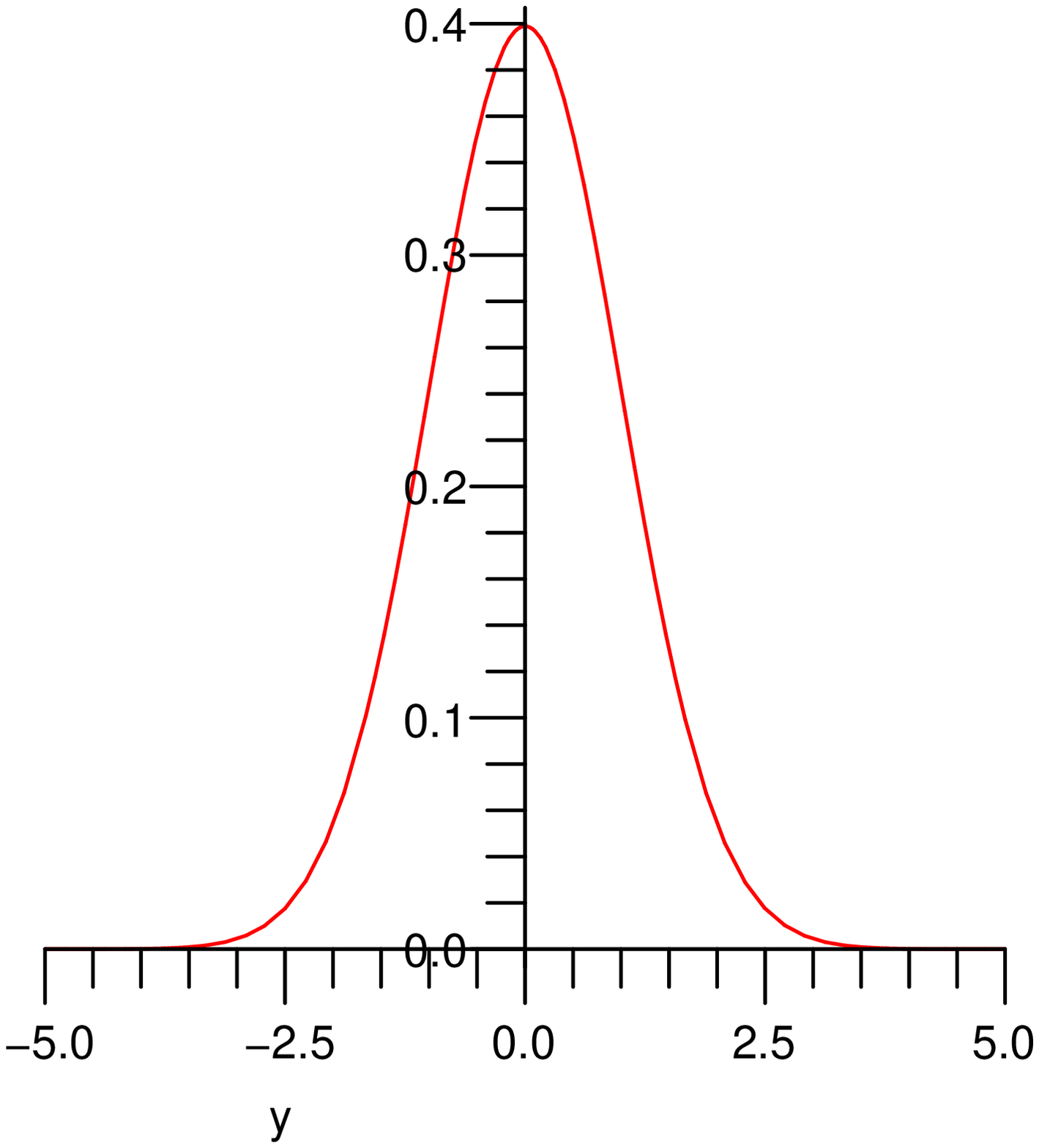, width=4.5cm}
\\[-2mm] {\footnotesize Figure 10: Probability density of the endpoint of the weighted path 
  for input $|\downarrow \, \rangle $.}
\end{minipage}\\
\end{tabular}\\[-1cm]
\end{center}
\begin{tabular}{l c r}
\begin{minipage}[t]{9cm}
\mbox{}\\[3mm]
In order to estimate $\sigma_z$, we use the pointer
of $\sigma_z$ illustrated here to the right. It is given by 
\begin{equation}\label{htilde}
\tilde{h}(x) = \frac{D_2}{x^2 + \delta} + D_3,
\end{equation}
with
$\delta = 2.701$, $D_2 = -21.649$ and $D_3 = 5.391$.
\end{minipage}
&
\hspace{0.2cm}
&
\begin{minipage}[t]{4.5cm}
\mbox{}\\*
  \epsfig{file=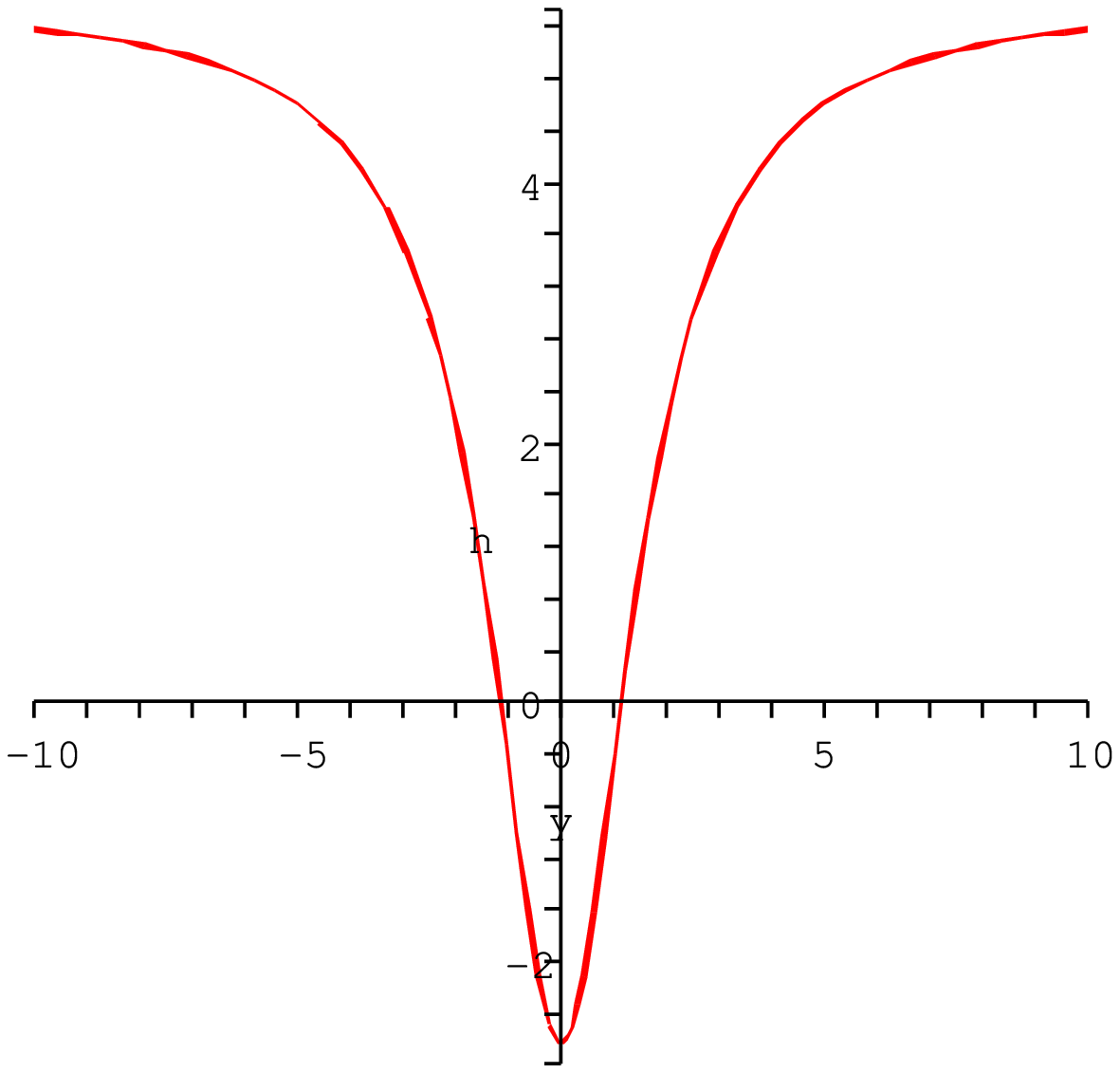, width=4cm}
  {\footnotesize Figure 11: Pointer for $\sigma_z$.}\\
  \mbox{}  
  \end{minipage}\\
\end{tabular}\\*
From formula \eqref{sofz}, we read off the probability distributions of this
pointer under the distributions of the endpoint of the weighted path illustrated in figures
8, 9 and 10. They are shown below:\\[-0.7cm]
\begin{center}
\begin{tabular}{r c l}
\begin{minipage}[t]{4.5 cm}
  \epsfig{file=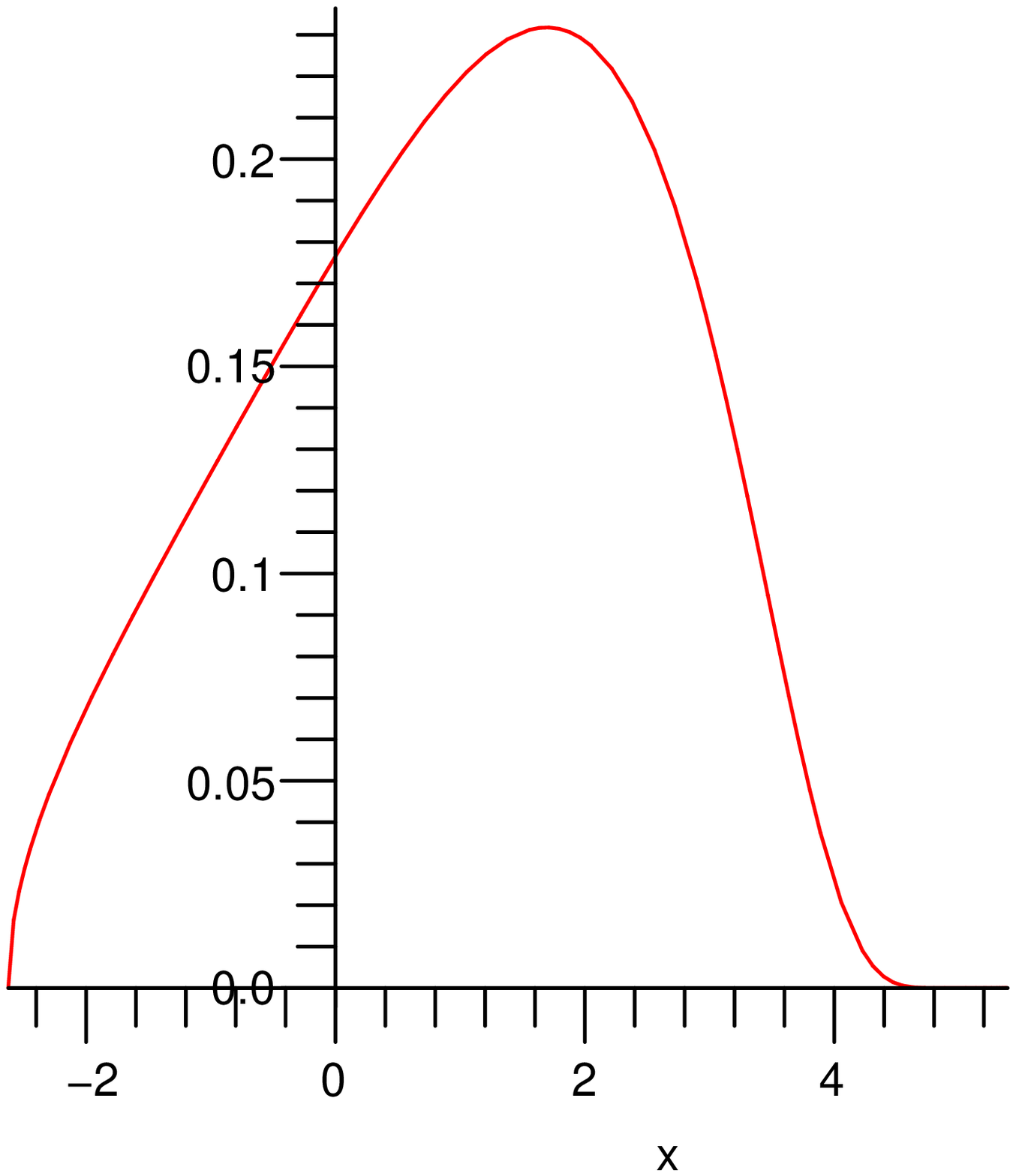, width=4.5cm}
 \\[-2mm] {\footnotesize Figure 12: Probability density of the $\sigma_z$-pointer  
  for input $|\uparrow \, \rangle $.}
\end{minipage}
&
\begin{minipage}[t]{4.5 cm}
  \epsfig{file=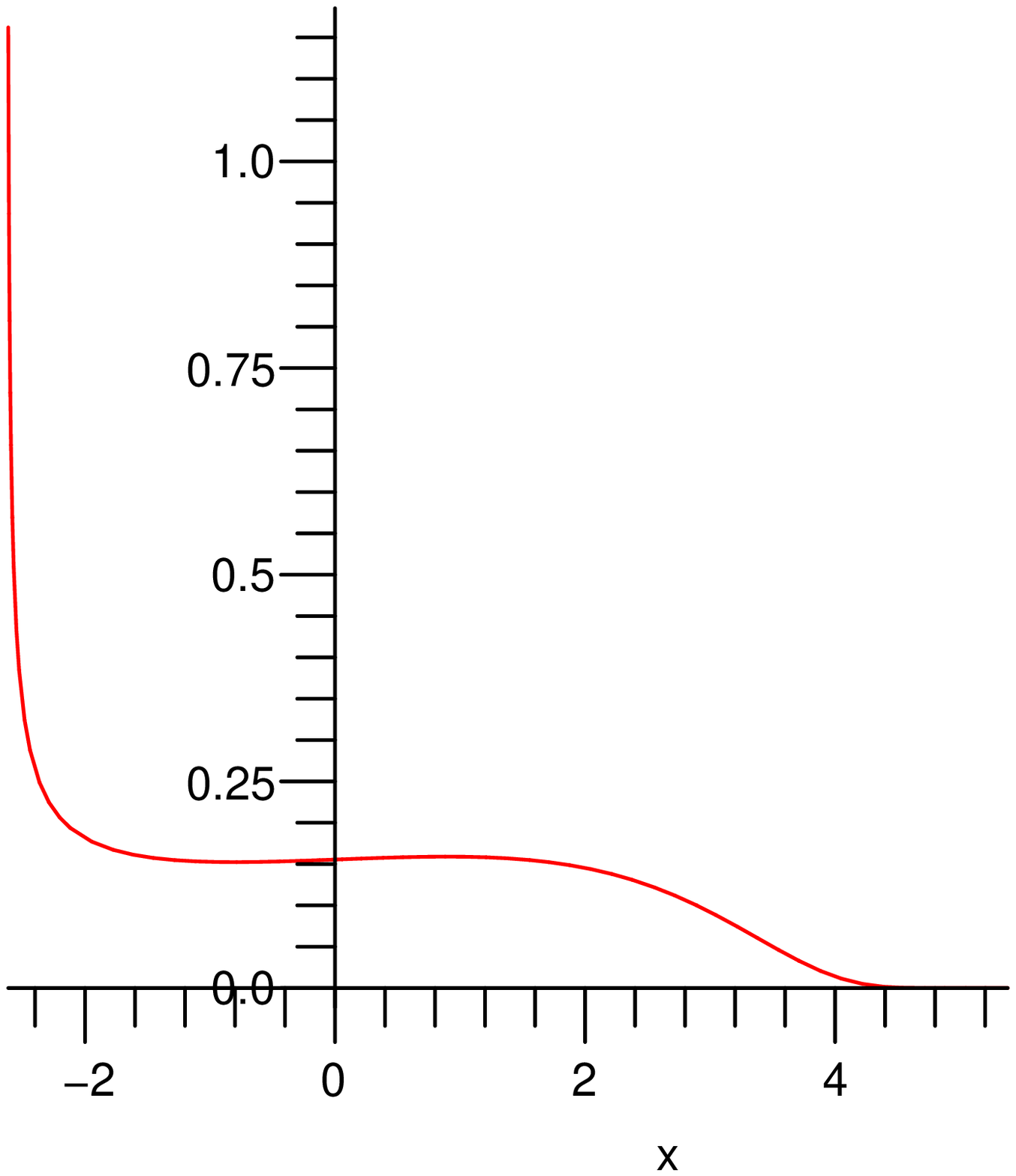, width=4.5cm}
\\[-2mm]  {\footnotesize Figure 13: Probability density of the $\sigma_z$-pointer  
  for input $\mbox{tr}$.}
\end{minipage}
&
\begin{minipage}[t]{4.5 cm}
  \epsfig{file=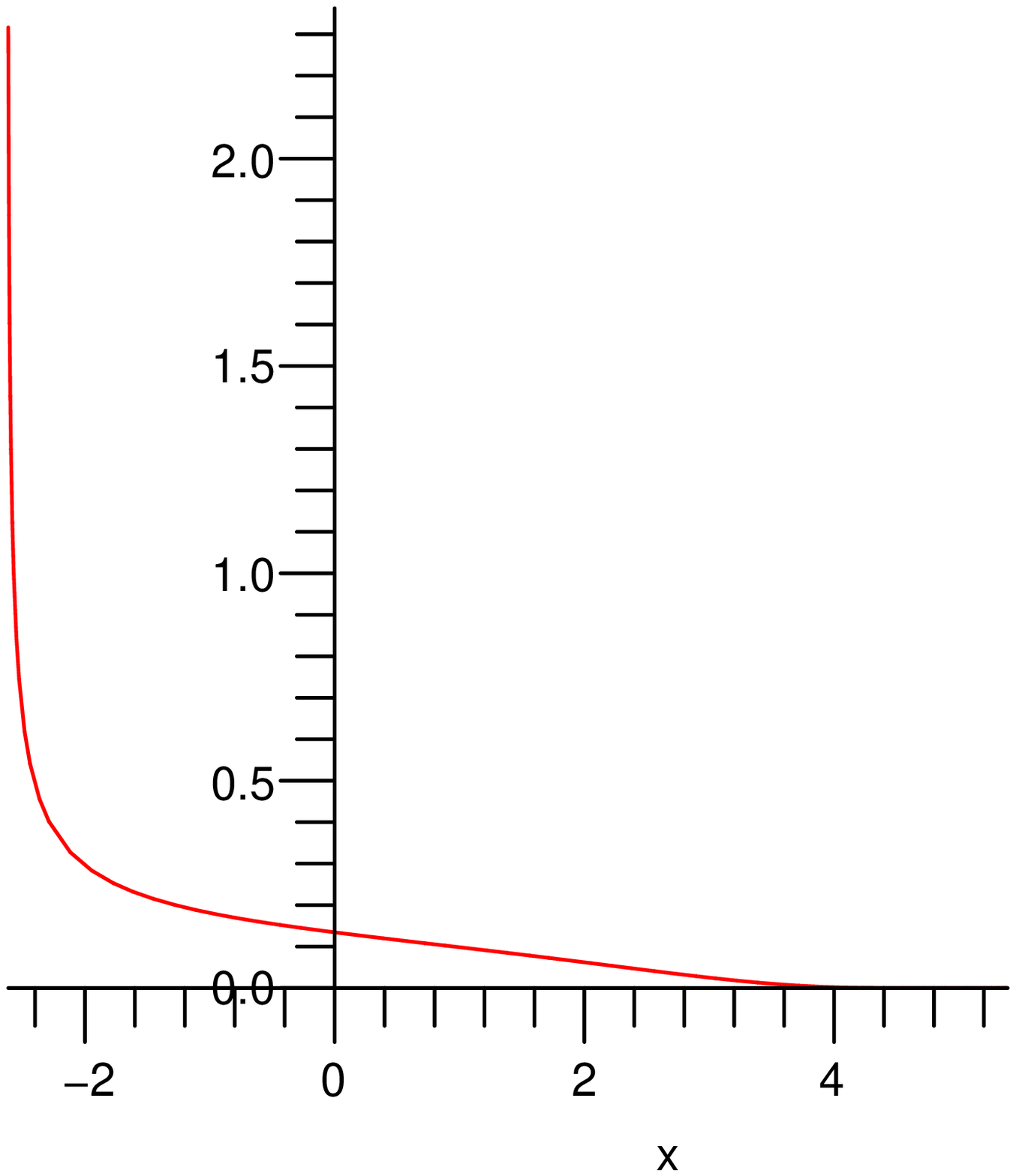, width=4.5cm}
 \\[-2mm] {\footnotesize Figure 14: Probability density of the $\sigma_z$-pointer 
  for input $|\downarrow \, \rangle $.}
\end{minipage}\\
\end{tabular}
\end{center}








\section{Discussion}\label{sec discussion}

In this paper, we have investigated homodyne detection of 
spontaneous decay of a two-level atom into the 
electromagnetic field.    
We have seen how the photocurrent, besides carrying information 
on $\sigma_x$ (which is immediate from the innovations
term in the filtering equation), also carries information 
on $\sigma_z$. 
Homodyne detection can thus be viewed as 
a joint measurement of the non-commuting observables $\sigma_x$ 
and $\sigma_z$ in the initial state of the qubit, and we have identified
the optimal pointers for this procedure in equations \eqref{hstar} and
\eqref{htilde}.

One particular feature of the pointers we constructed might seem counterintuitive at
first: they yield values outside $[-1,1]$ with nonzero probability.
This is a direct result of our requirement that the measurement be 
unbiased. Suppose, for example, that the input state is $| \uparrow \rangle$,
so that $\sigma_{z}$ has value 1. Since the photocurrent carries
information on $\sigma_{x}$ as well, its information on $\sigma_{z}$ is
certainly flawed, and will yield estimates $\sigma_{z} < 1$ at 
least some of the time. Unbiasedness then implies that also 
estimates $\sigma_{z} > 1$ must occur.   

On the other hand, an unbiased measurement will yield \emph{on average} 
the `true' value of
$\sigma_{z}$ for any possible input state. 
(Not just for the 3 possibilities
sketched on page \pageref{plotz}.) 
In repeated experiments, optimality of our pointers guarantees fast
convergence to these averages.  

Theorem \ref{thm joint} provides a theoretical bound for the quality of joint
measurement of $\sigma_{x}$ and $\sigma_{z}$. 
No conceivable measurement procedure can ever achieve 
$\sigma \tilde{\sigma} < 1$. 
It is now clear that
this bound cannot be met by way of homodyne detection:
a small part of the information extracted from the atom is simply lost
in this particular procedure.
Constructing the optimal pointers on the photocurrent does yield 
$\sigma \tilde{\sigma} = 1.056$ 
however, a figure much closer to the bound than the 4.437 provided by 
the na\"ive choice of \eqref{naive choice}.

\vspace{0.6 cm}

\textbf{\large Acknowledgment}\\
We thank M\u{a}d\u{a}lin Gu\c{t}\u{a} and Hans Maassen for 
interesting discussions. We thank John Gough for a critical 
reading of the text.
L.B.\ acknowledges support by the 
ARO under Grant DAAD19-03-1-0073.

\bibliography{ref}
\end{document}